\documentclass[numberedappendix, twocolumn, apj]{emulateapj}
\usepackage{graphicx}
\usepackage{natbib,amsmath}
\newcommand{\be}{\begin{equation}}
\newcommand{\ee}{\end{equation}}
\newcommand{\bs}{\boldsymbol}
\newcommand{\upa}{u_{\parallel}}
\newcommand{\uo}{u_{\parallel,0}}
\newcommand{\ug}{u_{\perp}}
\newcommand{\ugo}{u_{\perp,0}}

\begin{document}

\title{The Heating of Test Particles in Numerical Simulations of
  Alfv\'enic Turbulence}
\author{R\'emi Lehe\altaffilmark{1,2}, Ian J. Parrish\altaffilmark{2,3}, \& Eliot Quataert\altaffilmark{2}}
\altaffiltext{1}{D\'epartement de physique, \'Ecole Normale Sup\'erieure, 24 rue Lhomond, 75005 Paris, France}
\altaffiltext{2}{Astronomy Department \& Theoretical Astrophysics Center, 601 Campbell Hall, University of California, Berkeley, CA 94720}
\altaffiltext{3}{Einstein Fellow}
\shorttitle{TEST PARTICLE HEATING IN ALFV\'ENIC TURBULENCE}
\shortauthors{LEHE, PARRISH, \& QUATAERT}

\begin{abstract}

  We study the heating of charged test particles in three-dimensional
  numerical simulations of weakly compressible magnetohydrodynamic
  (MHD) turbulence (``Alfv\'enic turbulence''); these results are
  relevant to particle heating and acceleration in the solar wind,
  solar flares, accretion disks onto black holes, and other
  astrophysics and heliospheric environments.  The physics of particle
  heating depends on whether the gyrofrequency of a particle
  $\Omega_0$ is comparable to the frequency of a turbulent fluctuation
  $\omega$ that is resolved on the computational domain.  Particles
  with $\Omega_0 \sim \omega$ undergo strong {\it perpendicular}
  heating (relative to the local magnetic field) and {\it pitch angle
    scattering}.  By contrast, particles with $\Omega_0 \gg \omega$
  undergo strong {\it parallel} heating.  Simulations with a finite
  resistivity produce additional parallel heating due to parallel
  electric fields in small-scale current sheets.  Many of our results
  are consistent with linear theory predictions for the particle
  heating produced by the Alfv\'en and slow magnetosonic waves that
  make up Alfv\'enic turbulence.  However, in contrast to linear
  theory predictions, energy exchange is not dominated by discrete
  resonances between particles and waves; instead, the resonances are
  substantially ``broadened.''  We discuss the implications of our
  results for solar and astrophysics problems, in particular the
  thermodynamics of the near-Earth solar wind. This requires an
  extrapolation of our results to higher numerical resolution, because
  the dynamic range that can be simulated is far less than the true
  dynamic range between the proton cyclotron frequency and the
  outer-scale frequency of MHD turbulence.  We conclude that
  Alfv\'enic turbulence produces significant parallel heating via the
  interaction between particles and magnetic field compressions
  (``slow waves''). However, on scales above the proton Larmor radius
  Alfv\'enic turbulence does not produce significant perpendicular
  heating of protons or minor ions (this is consistent with linear
  theory, but inconsistent with previous claims from test particle
  simulations).  Instead, the Alfv\'en wave energy cascades to
  perpendicular scales below the proton Larmor radius, initiating a
  kinetic Alfv\'en wave cascade.

\end{abstract}

\keywords{turbulence -- MHD -- plasma -- particle acceleration --
  solar wind -- accretion disks}

\section{Introduction}

The heating and acceleration of particles by magnetohydrodynamic (MHD)
turbulence is believed to play a critical role in phenomena as diverse
as the solar wind and solar corona (e.g., \citealt{cranmer2005}),
accretion disks in active galactic nuclei (e.g.,
\citealt{Quataert:1999}), and the confinement and re-acceleration of
cosmic-rays in the Galaxy (e.g., \citealt{chandran2000,yan2002}).
This paper focuses on the physics of particle heating by {\it weakly
  compressible} MHD turbulence.  Weakly compressible MHD turbulence
consists of nonlinearly interacting Alfv\'en and slow magnetosonic
waves; since the Alfv\'en waves dominate the dynamics \citep{GS95}, we
will often refer to such turbulence as Alfv\'enic turbulence.
Measurements of magnetized turbulence in laboratory plasmas and in the
solar wind \citep{bale2005,sahraoui2009}, as well as analytic models
\citep{Shebalin:1983,Higdon:1984a,GS95} and numerical simulations
\citep{cho2000,maron2001} demonstrate that most of the energy in
Alfv\'enic turbulence cascades to small scales perpendicular to the
magnetic field.  This strongly influences how Alfv\'enic turbulence
couples to particles (e.g., \citealt{quataert98,leamon1999}).

The solar wind and solar corona provide a particularly rich source of
data on both the properties of MHD turbulence (e.g., the magnetic and
electric field power spectra) and on the possible thermodynamic
outcome of particle heating and acceleration by such turbulence (e.g.,
the proton, electron, and minor ion temperatures and/or distribution
functions).  A detailed review of these observational results and
their theoretical interpretation is not critical for this paper (see,
e.g., \citealt{marsch2006}).  However, one result that is particularly
germane is the fact that the proton distribution function is typically
anisotropic with respect to the local magnetic field: $T_\perp \simeq
0.9 \, T_\parallel$ in the solar wind at $\sim 1$ AU, although the
sign of the anisotropy depends on the wind speed
\citep{kasper2002,hellinger2006}.  In the fast solar wind, minor ions
such as O VI have $T_{\perp} \gg T_{\parallel}$, with
$T_{\perp}/T_{\parallel}$ increasing with the charge to mass ratio of
the ions \citep{marsch2006}.  An outstanding problem is whether these
measurements can be accounted for by heating by anisotropic Alfv\'enic
turbulence -- this is a particularly important question given the
growing body of evidence that the turbulent fluctuations in the solar
wind at $\sim 1$ AU {\it are} consistent with anisotropic Alfv\'enic
turbulence \citep{bale2005,sahraoui2009}.

The question of how Alfv\'enic turbulence heats and accelerates
particles is equally pressing in other astrophysical contexts.  For
example, in a particular class of models for accretion onto black
holes, Coulomb collisions between electrons and ions are too slow to
maintain thermal equilibrium, resulting in a two temperature plasma
with $T_i \ne T_e$ \citep{rees82}.  Weakly compressible MHD turbulence
is present in accretion disks because of the nonlinear saturation of
the magnetorotational instability \citep{balbus1998}.  In
two-temperature accretion disk models, the radiation from the
accreting plasma is produced largely by the (lighter) electrons, and
is thus sensitive to the details of how the turbulent energy is
thermalized as heat.

There is an extensive body of work studying the heating and
acceleration of (test) particles by plasma waves using ``quasilinear''
theory.  The key assumptions of this theory are that the orbits of the
particles are given by unperturbed helical motion around magnetic
field lines and that the plasma waves are long-lived; as a result,
energy exchange between waves and particles occurs only at discrete
resonances (e.g., \citealt{stix1992} and references therein; see \S
\ref{sec:th}).  To the extent that MHD turbulence can be approximated
as a superposition of roughly linear waves, the same resonances should
describe how magnetized turbulence couples to the underlying plasma.
This viewpoint dominates the literature on particle heating and
acceleration by MHD turbulence (e.g., \citealt{miller1995, quataert98,
  leamon1999, chandran2000, cranmer2003}).

In this paper, we go beyond linear theory by directly simulating the
orbits of charged {test particles} in three-dimensional numerical
simulations of weakly compressible MHD turbulence. Our numerical
simulations capture the physics of wave-particle interactions (e.g.,
cyclotron resonances) without making many of the simplifying
assumptions of standard linear theory.  In addition to wave-particle
resonances that can in principle operate throughout the inertial range
in a turbulent plasma, current sheets can form on small scales; these
current sheets represent an additional mechanism for particle heating.
In order to isolate the effects of current sheets, we calculate test
particle heating in simulations with and without an explicit
resistivity; we note up-front, however, that our MHD simulations
probably do not correctly capture how reconnection heats particles
(see \S \ref{sec:disc}).  Previous work on test particle heating in
numerical simulations of MHD turbulence \citep{dmitruck} concluded
that the preferential perpendicular heating of ions in the solar wind
summarized above could readily be accounted for.  We disagree with
this conclusion for reasons discussed in \S \ref{sec:disc}.

The test particle method used here has significant strengths, but also
some weaknesses.  Our grid-based MHD code can resolve a reasonably
large inertial range in a computationally efficient manner with test
particles spanning many orders of magnitude in gyrofrequency and/or
velocity, thus accurately capturing much of the relevant physics in a
single calculation.  The downside of the test particle approximation
is that efficient particle heating or acceleration would in reality
modify the turbulent cascade, an effect that is not captured by our
approach.  Fully electromagnetic particle-in-cell (PIC) calculations
of Alfv\'enic turbulence are not feasible with current computing
power.  An alternative approach, gyrokinetics, self-consistently
captures the physics of low frequency kinetic turbulence, and can be
used to study the combined problem of particle heating and its effect
on the turbulent cascade \citep{howes2008b}.  Gyrokinetic simulations
are, however, computationally intensive and order out higher frequency
dynamics such as the cyclotron resonances.  The test particle method
used here is thus a complementary approach that can be used to rapidly
and accurately explore the wave-particle interactions and reconnection
physics that occur on large-scales in MHD turbulence.

The remainder of this paper is organized as follows.  \S\ref{sec:th}
reviews the linear theory predictions for particle heating by
Alfv\'enic turbulence; these provide a useful framework for
interpreting our numerical results.  \S\ref{sec:numerics} outlines our
numerical methods, with some of the details given in the Appendix.  We
present the results of a fiducial simulation in \S\ref{sec:fiducial},
along with our interpretation of the resulting particle heating and
its relation to linear theory;
\S\S\ref{sec:resistivity}--\ref{sec:thermal} explore the effects of
varying the resistivity, magnetic field strength, and particle
distribution function.  Finally, in \S\ref{sec:disc} we summarize our
results and discuss their implications, focusing on the near-Earth
solar wind.

\section{Analytic expectations} \label{sec:th} 

Weakly compressible Alfv\'enic turbulence can be viewed as a
collection of nonlinearly interacting Alfv\'en and slow magnetosonic
waves \citep{GS95}.  We focus largely, but not entirely, on plasmas
with $\beta \gtrsim 1$, so that the linear dispersion relation of both
waves is given by $\omega \simeq |k_\parallel| v_A$, where $\omega$ is
the frequency of the wave, $k_\parallel$ is its wave-vector along the
local magnetic field, and $v_A$ is the local Alfv\'en speed.  In
quasilinear theory, these waves exchange energy with particles only at
discrete resonances, when (e.g., \citealt{stix1992})
\be \label{eq:res} \omega - k_\parallel u_\parallel = n \Omega \ee
where $u_\parallel$ is the particle's velocity along the magnetic
field, $\Omega$ is the particle's cyclotron frequency, and $n$ is an
integer. When equation (\ref{eq:res}) is satisfied for a linear wave,
there is phase coherence averaged over long timescales $\gg
\omega^{-1}$, and thus energy can be exchanged between the wave and
the particles.  In strong Alfv\'enic turbulence, however, nonlinear
interactions transfer energy from one scale to another on a timescale
comparable to the linear wave period ($\sim \omega^{-1}$).  It is thus
unclear whether standard quasilinear theory will adequately describe
the heating of particles by strong Alfv\'enic turbulence.  At a
minimum, we expect the discrete resonances in equation (\ref{eq:res})
-- which show up as delta functions in linear theory -- to be
significantly ``broadened'' due to the rapid decorrelation of the
phases of waves in strong turbulence (e.g., \citealt{chandran2000}).

The $n = 0$ resonance in equation (\ref{eq:res}) generally applies to
low frequency fluctuations having $\omega \ll \Omega$.  In this case
resonance occurs when $\omega = k_\parallel u_\parallel$, i.e., when
the parallel velocity of a particle is comparable to the parallel
phase speed of the wave, so that particles ``surf'' along with the
wave.  In the presence of such low frequency electromagnetic
fluctuations,
the magnetic moment of a particle, defined here as\footnote{In
  equation (\ref{eq:defmu}), we have dropped the conventional factor
  of $m/2$, since the masses of the particles we consider are
  arbitrary.}  \be\label{eq:defmu} \mu = \frac{\ug^2}{B} \ee is an
adiabatic invariant and remains approximately constant ($B = |\bs{B}|$
is the magnitude of the magnetic field strength and $\ug$ is the
gyration/perpendicular velocity).

It is straightforward to show that in the limit $\omega \ll \Omega$,
the equation of motion for a charged particle simplifies to
\citep{achterberg1981} \be \label{eq:para} \frac{d \upa}{dt} =
\frac{q}{m} E_{\parallel} - \mu \nabla_{\parallel} B \ee and
\be \label{eq:perp} \frac{d \ug}{dt} = \frac{\ug}{2
  B}\left(\frac{\partial B}{\partial t} + \upa \nabla_{\parallel} B
\right) \ee where $E_\parallel$ is the parallel electric field and
$\nabla_\parallel$ is the gradient along the local magnetic field;
note that equation (\ref{eq:perp}) refers to the magnitude of the
perpendicular velocity and so does not contain the cyclotron motion.

Equation (\ref{eq:para}) demonstrates that the $n = 0$ resonance
contains two physically distinct heating mechanisms, that due to
parallel electric fields (Landau damping) and that due to the $\mu
\nabla B$ force (transit-time damping); for simplicity, however, we
will refer to this low-frequency resonant interaction as the ``Landau
resonance.''  In MHD, linear Alfv\'en waves have $\delta E_\parallel,
\delta B = 0$ and thus cannot interact with particles via the Landau
resonance.  Alfv\'en waves do develop $\delta E_\parallel \ne 0$
and/or $\delta B \ne 0$ on small scales where MHD breaks down (e.g.,
\citealt{quataert98}), but this is not captured by our turbulence
simulations which utilize MHD.

Unlike the Alfv\'en wave, the slow wave has $\delta B \ne 0$ in MHD;
analytic studies of particle heating by anisotropic Alfv\'enic
turbulence have thus predicted that the slow wave should produce
significant heating via the Landau resonance (e.g.,
\citealt{blackman99}).  In linear theory, this is predicted to be
entirely {\it parallel} heating, i.e., it should increase the parallel
temperature, but not the perpendicular temperature, of the particles.
This follows directly from equation \eqref{eq:perp} by considering a
single Fourier component with $\delta B \propto e^{-i \omega t + i
  {\bf k \cdot r}}$, in which case $d \ug/dt \propto [\omega -
k_\parallel u_\parallel]$.  By construction, however, $\omega =
k_\parallel u_\parallel$ at the Landau resonance and so $d \ug/dt =
0$.

If $\omega \sim \Omega$, i.e., if the fields vary significantly on the
timescale of a particle's cyclotron motion, strong ``cyclotron
resonance'' may occur; this corresponds to $n = \pm 1$ in equation
(\ref{eq:res}).\footnote{We do not consider higher $n$ resonances
  here.}  Cyclotron resonance occurs when the perpendicular electric
force due to a wave remains in phase with the rotating cyclotron
motion of a particle.  Because $\omega \sim \Omega$, adiabatic
invariance of $\mu$ is violated and cyclotron resonance leads to
strong pitch angle scattering and perpendicular heating.

The Alfv\'en wave component of anisotropic MHD turbulence is a
transverse linearly polarized wave that can undergo cyclotron
resonance with both negatively and positively charged particles.  By
contrast, anisotropic slow waves are largely longitudinal and thus do
not undergo strong cyclotron resonance.  To estimate the conditions
required for cyclotron resonance in anisotropic MHD turbulence, we use
Goldreich \& Sridhar's (1995) critical balance conjecture that relates
the typical parallel ($k_\parallel$) and perpendicular ($k_\perp$)
wavevectors of turbulent fluctuations: $k_\parallel \simeq
k_\perp^{2/3} k_{min}^{1/3}$, where $k_{min}^{-1}$ is the outer scale
of the turbulence.  This scale-dependent anisotropy of incompressible
MHD turbulence has been confirmed in a number of numerical studies
\citep{cho2000,maron2001}.  Together with the dispersion relation for
Alfv\'en waves in MHD, $\omega = |k_\parallel| v_A$, this implies that
cyclotron resonance occurs when \be \label{eq:resonance}
k_{\perp}^{2/3} k_{min}^{1/3} \left(v_{A} -
  u_{\parallel}\frac{k_\parallel}{|k_\parallel|}\right) = \pm \Omega.
\ee Equation (\ref{eq:resonance}) will be useful for interpreting some
of the numerical results described later in this paper.

To summarize the above discussion, quasilinear theory predicts that
anisotropic Alfv\'enic turbulence resonantly couples to particles in
two ways: (1) parallel $\mu \nabla_\parallel B$ heating of particles
by Landau-resonant slow modes, (2) pitch-angle scattering and
perpendicular heating of particles by cyclotron-resonant Alfv\'en
waves.  Quasilinear theory does not account for possible heating of
particles at current sheets, but this will naturally be a feature of
our resistive MHD simulations, in addition to the wave-particle
resonances reviewed here.

\section{Numerical methods}
\label{sec:numerics}

Our simulations focus on collisionless test particles evolving in isothermal subsonic MHD turbulence. Our approach involves two distinct computational phases. First, the macroscopic fields are evolved according to the MHD equations. Then the test particles' positions and velocities are updated, using the macroscopic fields to compute the appropriate Lorentz force.

\subsection{The MHD Integrator}
\label{sec:MHD}

To compute the macroscopic fields, we use the Athena MHD code
\citep{gs05,sg08}.  Calculations are done on a uniform Cartesian grid
with periodic boundary conditions. Our fiducial resolution features
256$^3$ zones, but some high resolution calculations are done at
512$^3$. We initialize the grid with uniform fields. A background
magnetic field is set along the $x$-axis, with a magnitude $B_0$
determined by the chosen value of $\beta = \rho c_s^2 /[B_0^2/8\pi]
$, where $\rho$ is the fluid density and $c_s$ is its sound speed. The
fluid velocity is initially zero. Kinetic energy is then steadily
injected into the box, and the turbulence progressively grows.

We drive the turbulence with a method similar to that of \citet{ls09}.
At each timestep, we generate a velocity perturbation with a random
amplitude in Fourier space, which is then added to the fluid's
velocity. The Fourier components of this velocity perturbation are
non-zero only for $0 < k \leq 4 \times \frac{2 \pi}{L}$ (where $L$ is
the size of the simulation box), so that we are exclusively driving on
large scales. We project each Fourier component perpendicular to
$\bs{k}$, so as to make it divergence-free. This avoids explicitly
driving compressible modes. Finally, we ensure that the energy
injection rate ($\dot{E}$) is constant at each timestep; it is given
roughly by $\dot E \simeq L^2 \rho \delta v^3$, where $\delta v$ is
the magnitude of the turbulent velocity once the turbulence has
saturated.

After each random perturbation, the fields evolve according to the conservative equations of isothermal MHD, either ideal or resistive, depending on the value of $\eta $:

\be \frac{\partial \rho}{\partial t} + \bs{\nabla}\cdot(\rho \bs{v})=0 \ee
\be \frac{\partial \rho \bs{v}}{\partial t} + \bs{\nabla} \cdot \left[ \rho \bs{v}\bs{v} - \frac{\bs{B}\bs{B}}{4 \pi} + \left(P + \frac{\bs{B}^2}{8 \pi} \right) \bs{I}\right] = \bs{0} \ee
\be \frac{\partial \bs{B}}{\partial t} = \bs{\nabla} \times ( \bs{v} \times \bs{B} ) + \frac{\eta c}{4\pi} \bs{\nabla}^2 \bs{B} \ee

In our simulations, these equations are implemented by evaluating the conservative fluxes with the HLLD Riemann solver \citep{mk05} and then evolving the fields using a Van Leer integrator \citep{sg09}. The constrained-Transport (CT) algorithm ensures that the magnetic field remains divergence-free. In the case of resistive MHD ($\eta \neq 0$), the evolution of the fields is operator-split: at each timestep, we first evolve the fields according to the ideal equations, and then apply the magnetic diffusion operator in a way that is consistent with CT.

Our simulations are run with $ L = \rho = c_{s} = 1.0 $, but are
actually scale-free. Our results can thus be transformed to any other
physical parameters by scaling them with an appropriate combination of
the new $L$, $\rho$ and $c_{s}$. The behavior of the turbulence is
controlled by two dimensionless numbers: the energy injection rate
$\dot{E}$ (in units of $\rho L^2c_s^3$) and the dimensionless magnetic
field strength (e.g., $\beta$ or $v_A/c_s$).

Our choices of $\dot{E}$ and $\beta$ produce both low sonic and low
Alfv\'enic Mach numbers ($\delta v/c_s < 1$ , $\delta v/v_A <
1$). Therefore, our simulations lie in the regime of weakly
compressible Alfv\'enic turbulence. As the simulation proceeds, the
energy cascades from low wavenumber, where it is injected, to higher
\emph{perpendicular} wavenumber, thus forming an inertial range. The
cascading energy is eventually damped by physical and numerical
dissipation at very high wavenumber. After some time (a few $L/\delta
v$), the energy dissipation rate equals the injection rate, and the
amplitude of the turbulence saturates. Diagnostics of the power
spectra in the saturated state agree reasonably well with the
Goldreich-Sridhar predictions and with previous numerical simulations;
e.g., our results are consistent with a $k_{\perp}^{-10/3}$ spectrum
of the energy density and most of the energy is contained at
perpendicular wavenumbers $k_\perp \gg k_\parallel$, consistent with
the critical balance conjecture.

\subsection{The Particle Integrator}
\label{sec:particle}
\begin{figure} 
\centering
\includegraphics[scale=0.45]{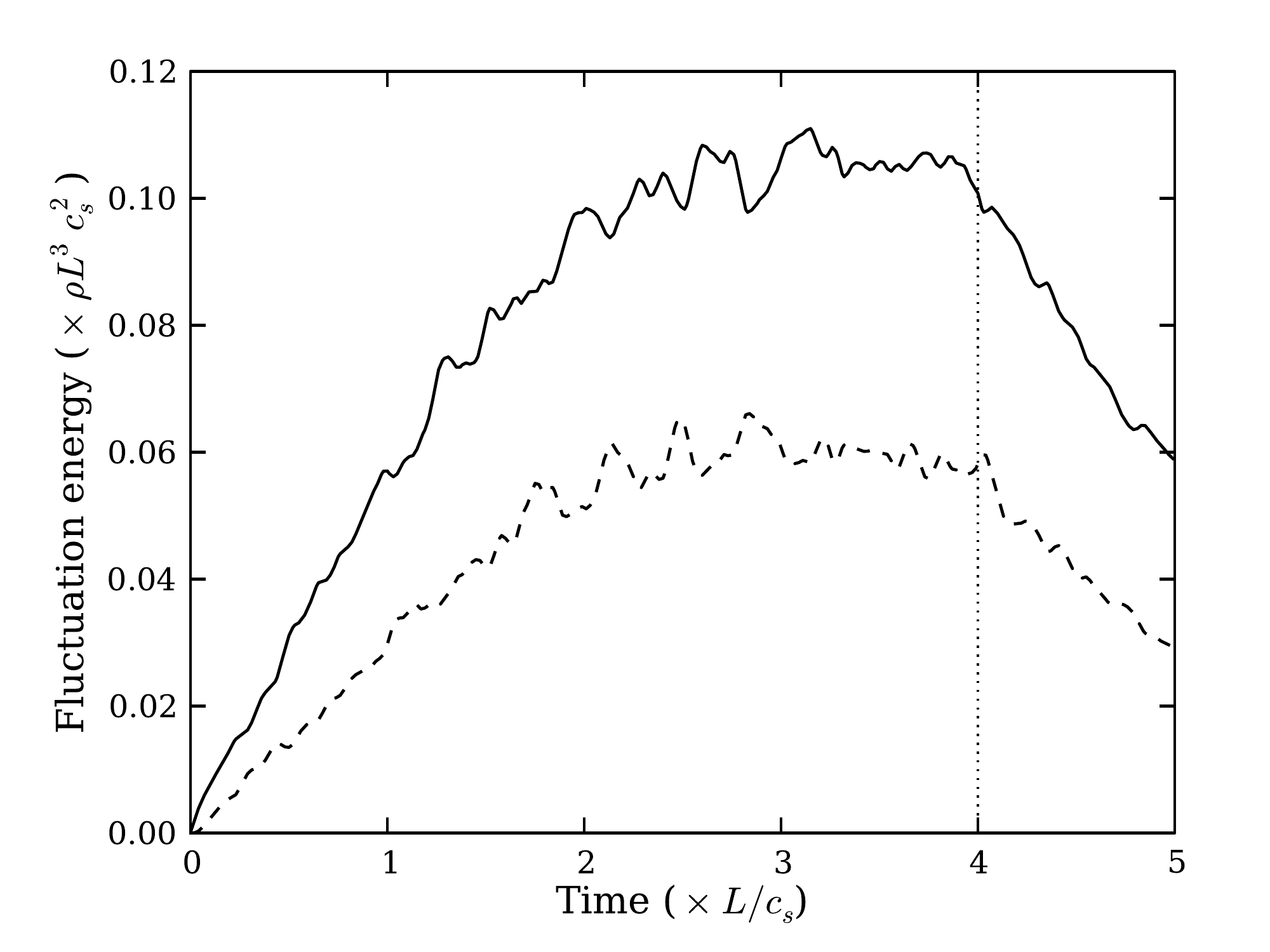}
\caption[Turbulent energy vs. time]{Energy in the turbulent
  fluctuations versus time. The solid and dashed lines correspond to
  the total kinetic energy and magnetic energy, respectively. The
  vertical dotted line marks the time at which the particle
  integration begins.  After that time the turbulence decays, to avoid
  spurious particle heating due to the temporal driving (see Appendix
  \ref{ap:driving}).}
\label{fig:hst}
\end{figure}

Once the saturated state of the turbulence is reached (at $t = 4\,
L/c_s$), we turn on the particle integrator. From then on, the
particles and macroscopic fields are evolved simultaneously. The
periodic boundaries of the simulation box also apply to the
particles. We found that the driving of the turbulence could introduce
spurious acceleration of the particles (see \S \ref{ap:driving} of the
Appendix).  In order to avoid this we stop driving the turbulence
after the particles have been injected, and let the turbulence
progressively decay. The particle integrator usually runs for a
physical time of $1.0 \, L/c_s$. Figure \ref{fig:hst} shows the
evolution of the turbulence during a typical run. The energy first
builds up and saturates, and then decays as we inject the particles
and turn off the driving.

Our collisionless charged particles evolve according to: 
\be \label{eq:motion} \frac{d\bs{u}}{dt} = \frac{q}{m}\bs{E} +
\frac{q}{mc}\bs{u}\times\bs{B} \ee 
where $\bs{u}$ is the particle's velocity and the remaining symbols have their standard meanings. In the particle integrator, the fields $\bs{E}$ and $\bs{B}$ are interpolated in space and time from their known values on the grid to the particle's position. We obtain $\bs{E}$ from $\bs{v}$ and $\bs{B}$ on the grid by using:
\be 
\bs{E} = -\frac{1}{c}\bs{v}\times\bs{B} + \frac{\eta}{4\pi}\bs{\nabla}\times\bs{B} 
\ee

The particle's charge-to-mass ratio ($q/m$ in equation
eq. [\ref{eq:motion}]) is a free parameter. Our simulations
simultaneously evolve different types of particles, with
charge-to-mass ratios spanning several orders of magnitude. To make
the results easier to interpret, we convert the charge-to-mass ratio
to a normalized gyrofrequency: $\Omega_0 = qB_0/mc$. The actual
gyrofrequency $\Omega$ of a particle is close to $\Omega_0$, but may
differ due to local turbulent fluctuations in the $\bs{B}$ field.

In our runs, particles all have positive charge. Since anisotropic
Alfv\'enic turbulence consists of slow waves and linearly polarized
Alfv\'en waves, the sense of the gyration (due to the sign of the
charge) should make no difference to the particle heating and
acceleration. To confirm this, we compared simulations with negatively
charged particles to those with positively charged particles, and
observed no significant differences.

\subsubsection{Integration method} \label{sec:integration}

We first implemented the 4$^{th}$-order Runge-Kutta method and found
that it was not suitable for this specific problem. We performed tests
using a constant, uniform magnetic field (instead of the turbulent
fields). Those tests showed a small but steady decrease in the
gyroradius and the energy of the particles since the sign of the error is always in the same direction. With, e.g., 10 time steps
per gyration, the energy decreased by 1 percent per gyration. This
numerical bias is particularly dangerous given our goal of studying
particle heating.

We thus replaced the Runge-Kutta method by the following implicit
leap-frog numerical scheme, where positions are defined at integer
time and velocities at half integer time:
\begin{align} \label{eq:scheme}
\frac{\bs{u}^{n+1/2} - \bs{u}^{n-1/2}}{\Delta t} & = \frac{q}{m}\bs{E}^{n} \nonumber \\
& + \frac{q}{mc} \frac{ (\bs{u}^{n+1/2} +\bs{u}^{n-1/2} ) }{2} \times \bs{B}^{n} \\
\frac{\bs{x}^{n+1} - \bs{x}^{n}}{\Delta t} & = \bs{u}^{n+1/2}
\end{align}
where a superscript $n$ indicates that the variable is defined at time level $n$. 

The specific implementation of equation \eqref{eq:scheme} that we use is the Boris particle pusher \citep{boris}. Unlike the Runge-Kutta method, this scheme is symmetric in time and symplectic, thus conserving energy and adiabatic invariants to machine precision. Another advantage of this method is that the fields only need to be interpolated once per timestep, which makes it considerably faster than the 4$^{th}$-order Runge-Kutta method. 

We choose the particle timestep $\Delta t$ so as to properly resolve
the gyration. In order to also ensure that the particle timestep is
smaller than the MHD timestep, we use: \be \Delta t = \mathrm{min}
\left( \frac{1}{40} \frac{2\pi}{\Omega_0}, \frac{1}{10} \Delta t_{\mathrm{MHD}}
\right) \ee which produces at least 40 steps per gyration. Tests using
simplified field configurations confirmed the accuracy of this
method. Some of our tests are described in \S \ref{ap:tests} of the
Appendix.

\subsubsection{Interpolation}\label{sec:interpolation}

The $\bs{B}$ and $\bs{E}$ fields are interpolated from their values on
the grid and at the discrete MHD timesteps to the particle's
position. We use the Triangular Shaped Cloud (TSC) interpolation
method \citep{tsc} in space \emph{and} time. More details on the
algorithm can be found in \S \ref{ap:interpolation} of the Appendix.

In ideal MHD, $E_{\parallel} = 0$. Although this is indeed true on the
grid, the \emph{interpolated} $\bs{E}$ may have a non-zero component
along the \emph{interpolated} $\bs{B}$. Because the particles' motion
parallel to the magnetic field is unimpeded, this component, although small, can produce significant acceleration. In order to
avoid non-physical parallel energization, we compute
$\bs{E}\cdot\bs{B}$ on the grid (this is zero in ideal MHD, but
non-zero in resistive MHD) and interpolate it to the particle's
position. We then modify the component of $\overline{\bs{E}}$ parallel
to $\overline{\bs{B}}$: \be \widetilde{\bs{E}} = \overline{\bs{E}} +
\left( \overline{(\bs{E} \cdot \bs{B})} - \overline{\bs{E}} \cdot
  \overline{\bs{B}} \right) \frac{\overline{\bs{B}} }{
  ||\overline{\bs{B}}||^2} \ee where the overline symbolizes
interpolation and $\widetilde{\bs{E}}$ is the new electric field.
This ensures that: \be \widetilde{\bs{E}}\cdot \overline{\bs{B}} =
\overline{(\bs{E}\cdot \bs{B})} \ee i.e., that the parallel component
of $\widetilde{\bs{E}}$ is correct.

\subsection{Measurement and Initialization of the Particles}\label{sec:init}

We initialize the particles uniformly over the entire computational
domain. We then give them a parallel velocity $\upa$ and a gyration
velocity $\ug$ drawn from the distributions discussed below.

Since the motion of a particle is in fact the superposition of a rapid
gyration and a slowly varying drift, calculating the gyration velocity
$\ug$ requires knowledge of the drift velocity: $\ug = ||
\bs{u}_{\perp,tot} - \bs{u}_{d} ||$. ($\bs{u}_{\perp,tot}$ and
$\bs{u}_d$ are the total perpendicular velocity of the particle and
its drift velocity, respectively.)  We take into account the dominant
$\bs{E}\times\bs{B}$ drift but neglect the other drifts ($\nabla B$,
curvature drift, polarization drift, ...), which are small for the
particles considered here. For simplicity, we also neglect the
$\bs{E}\times\bs{B}$ drift due to the \emph{resistive} part of
$\bs{E}$,\footnote{This drift is proportional to $\eta c/4 \pi$, and
  is always small.} and adopt the following definition: \be \ug = ||
\bs{u}_{\perp,tot} - \bs{v}_{\perp} || \ee where $\bs{v}_{\perp}$ is
the perpendicular velocity of the fluid.

We use two different kinds of initial velocity distributions. For some
of our calculations, our aim is to isolate the statistical effect of
the turbulence on a certain type of particle. In this case, all of the
particles have the same initial parallel velocity and magnetic moment:
\be \label{eq:init} f_{0}(\bs{x}, \bs{u}) \propto \delta \left( \upa -
  \uo \right) \delta \left( \ug - \sqrt{ \mu_{0} B(\bs{x}) } \right)
\ee where $\uo$ and $\mu_{0}$ specify the initial particle properties.
We give the particles the same magnetic moment, not the same gyration
velocity, because analytic theory predicts that $\mu$ should be
conserved for particles with high $\Omega_0$ (\S \ref{sec:th}).  With
this initialization we can easily check whether the interaction with
the turbulence indeed conserves $\mu$.  This test has been extensively
performed while developing our methods.

By contrast, in other calculations, we study the integrated heating
over a ``realistic'' distribution of particles. Here, we choose to
assign the velocities according to an isotropic Maxwell-Boltzmann
distribution: \be \label{eq:thermal} f_{MB}(\bs{x}, \bs{u}) \propto
\exp\left[ - \frac{\left( \ug^2 + \upa^2 \right) }{2 c_{s}^2} \right]
\ee In all cases the particles are split into $N$ groups, where in
each group, particles all have the same physical properties such as
charge-to-mass ratio (i.e., $\Omega_0$) or initial velocity.  Each of
the N groups contains particles with different values of the physical
property being studied in the particular simulation (e.g., $\Omega_0$
or velocity).

\section{Results}

In the following sections we describe the interaction between the
turbulence and particles with different gyrofrequencies $\Omega_0$ and
velocities (\S \ref{sec:fiducial}).  We also compare simulations with
and without explicit resistivity (\S \ref{sec:resistivity}),
simulations with different background magnetic field strengths (\S
\ref{sec:beta}), and simulations with different grid resolution (\S
\ref{sec:resolution}). Finally, we study the net effect of the
turbulence on a thermal distribution of particles (\S
\ref{sec:thermal}).  To help guide the reader's intuition, we note
that our fiducial calculations (see Table \ref{tab:fiducial}) have
$\beta \sim 1$ and turbulence with outer-scale velocities $\delta v
\sim 0.5 c_s$; the outer-scale eddy turnover time is $\sim 2 L/c_s$.
We consider particles with a range of velocities $\sim 0.1-10 \,
c_s$. High velocity particles ($\gg c_s$) transit the box many times
in the course of the simulation; doing so, they may unphysically
sample a similar realization of the turbulent fluctuations.  We thus
focus on lower velocity particles where the periodic box boundary
conditions we employ are more reliable.  This corresponds primarily to
the velocities of protons or minor ions, rather than electrons, which
have typical velocities $\sim 40 \, c_s$ for $T_e \sim T_p$.

The range of gyrofrequencies we consider is $\sim 10-10^4 \,
c_s/L$. For a resolution of 256$^3$, the highest linear frequency on
the computational domain is $\omega_{max} \simeq k_{\parallel, max}
v_A \sim 350 \, c_s/L$ where we have used the critical balance
conjecture (\S \ref{sec:th}) to estimate that for anisotropic
Alfv\'enic turbulence, $k_{\parallel, max} \simeq k_{max}^{2/3} \,
k_{min}^{1/3}$ ($k_{max} \simeq \pi/\Delta x$ is the highest
wavenumber given a resolution $\Delta x$ and $k_{min} \simeq 2
\pi/(L/4)$ is the wavenumber at which the turbulence is driven).  Thus
the particle gyrofrequencies we consider range from particles with
gyrofrequencies comparable to those of resolved turbulent fluctuations
to particles with $\Omega_o \gg \omega_{max}$.  From linear theory,
the latter are expected to have primarily parallel heating because of
the conservation of $\mu$ (\S \ref{sec:th}).

\renewcommand{\arraystretch}{1.5}

\begin{deluxetable}{p{4.5cm} p{2.5cm}} \label{tab:particles}
\tablecolumns{2}
\tablecaption{Properties of the Fiducial Calculation\label{tab:fiducial}} 
\tablewidth{0pt}
\tablehead{
Parameter & Value }
\startdata
$\beta$ & $1.0$ \\
$\dot{E}$ & $0.1\,\rho L^2 c_{s}^3$ \tablenotemark{a}\\
$\eta$ & 0 (ideal MHD) \\
Resolution & $256^3$ zones \\
Integration time\tablenotemark{b} & $1.0\, L/c_s$ \\
Number of particles\tablenotemark{c} & $512 \times 10^3$  
%$\uo$ & $1.0\,c_s $ \\
%$\mu_0$ & $1.0 \, c_{s}^2 / B_0$\tablenotemark{b,c}\\
%$\Omega_{min}$ & $ 5.0 \; c_{s}/L$ \\
%$\Omega_{max}$ & $ 2.0 \times 10^4 \; c_s/L$ \\
%Number of groups with a given $\Omega_0$ &100 \\
%Number of particles per group & 5,120 \\
%\hline
\vspace{0.1cm}
\enddata
\tablenotetext{a}{This produces a sonic Mach number of $\approx 0.46$.  Note that we stop driving the turbulence once the particle integration begins (see Fig. \ref{fig:hst} \& \S \ref{sec:particle}).}
\tablenotetext{b}{The time interval between the initialization of the particles and the end of the simulation.}
\tablenotetext{c}{The particles are
  initialized with a delta function in $u_{\parallel,0}$ and $\mu_0$
  (eq. [\ref{eq:init}]); we consider a range of $\Omega_0 \simeq 10-10^4 \, c_s/L$.}
%\tablenotetext{b}{$B_0 = \sqrt{8\pi\rho/\beta} \, c_s$}
%\tablenotetext{c}{This value yields an average initial $\ug$ of $1.0\, c_s$.}

\end{deluxetable}

\renewcommand{\arraystretch}{1.5}

\subsection{Fiducial Results}\label{sec:fiducial}

\begin{figure} 
\centering
\includegraphics[scale=0.45]{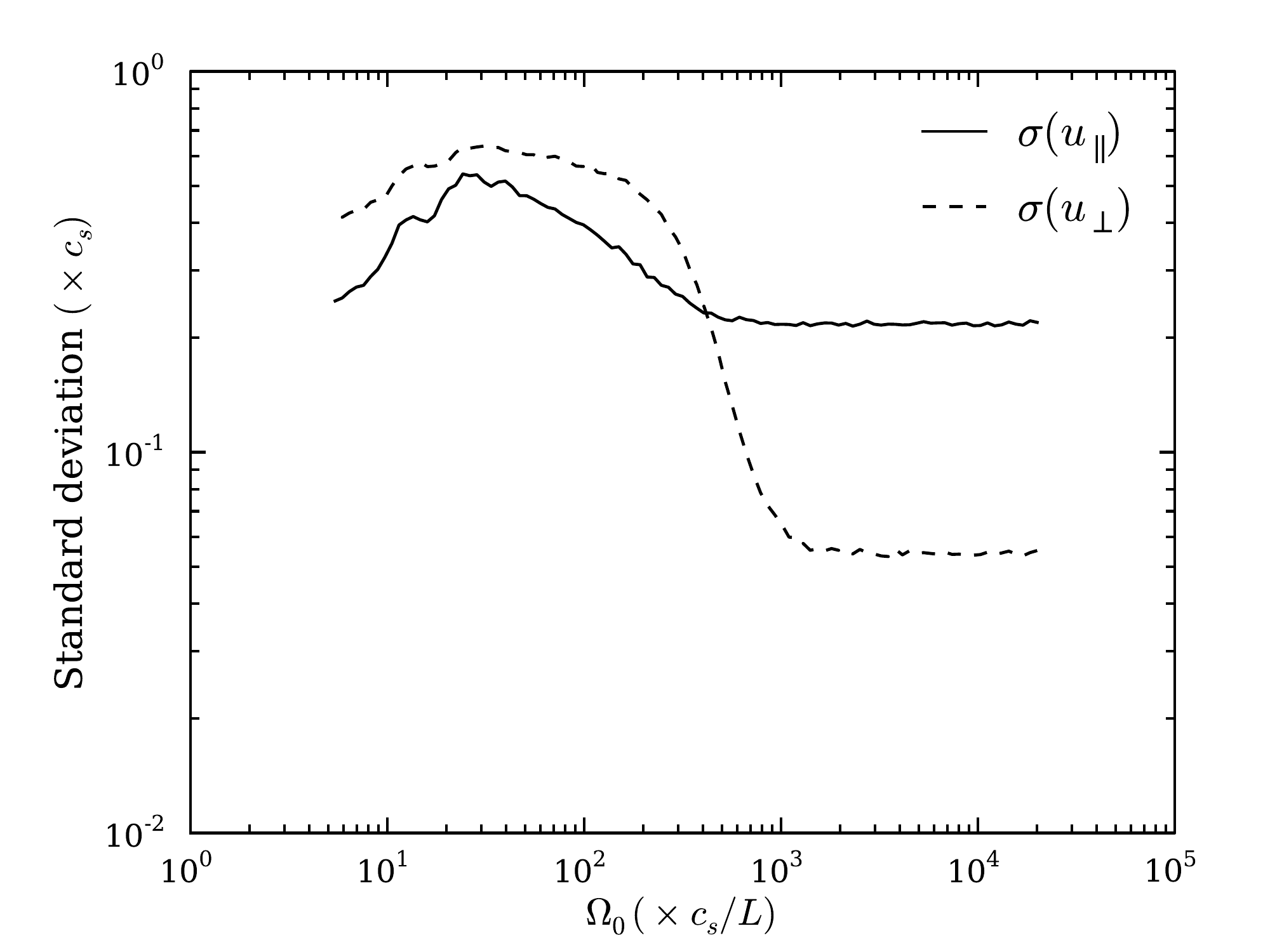}
\caption{Standard deviation of the parallel $\upa$ (solid line) and
  perpendicular $\ug$ (dashed line) velocities at the end of the
  simulation, as a function of the particles' gyrofrequency
  $\Omega_0$; the particles initially have delta functions in velocity
  with $\uo = 1.0 \, c_s$ and $\mu_0 = 1.0 \, c_s^2/B_0$ ($\ugo \simeq
  1.0 \, c_s$).  The strong diffusion for low $\Omega_0$ particles is
  consistent with linear theory predictions for cyclotron resonance;
  $\sigma(\upa) \gg \sigma(\ug)$ for high $\Omega_0$ particles is
  consistent with linear theory predictions for $\mu \nabla_\parallel
  B$ acceleration at the Landau resonance (see \S \ref{sec:fiducial}
  for details).  The parameters of this calculation are summarized in
  Table \ref{tab:fiducial}.}
\label{fig:dispersion}
\end{figure} 

\begin{figure} 
\centering
\includegraphics[scale=0.45]{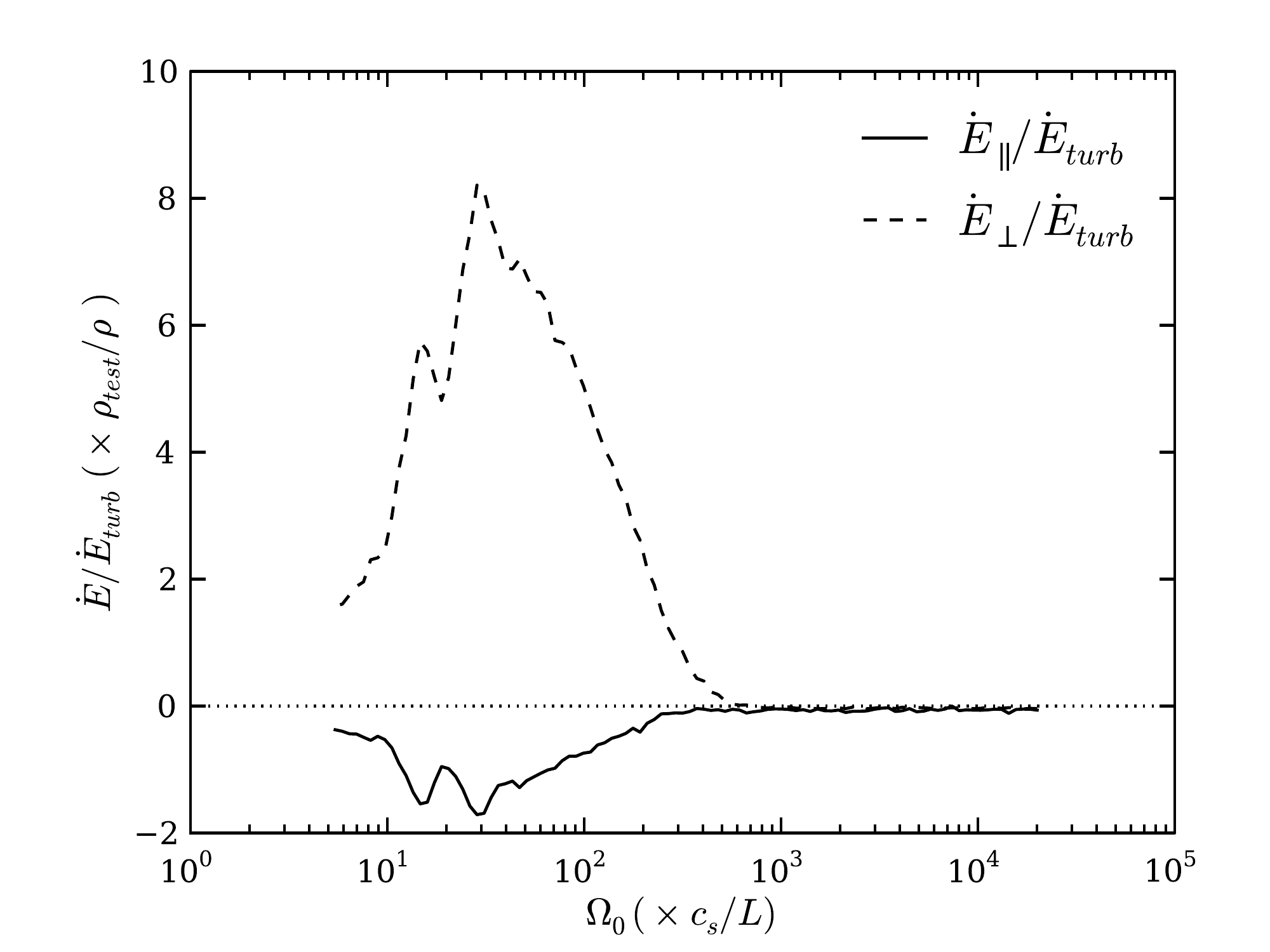}
\caption{Parallel (solid) and perpendicular (dashed) heating rates as
  a function of the particles' gyrofrequency $\Omega_0$; the particles
  initially have delta functions in velocity with $\uo = 1.0 \, c_s$
  and $\mu_0 = 1.0 \, c_s^2/B_0$ ($\ugo \simeq 1.0 \, c_s$). The
  heating rates are normalized by $\dot{E}_{turb}$, the energy
  dissipation rate in the turbulence. The heating rates are averaged
  over the full $1.0 L/c_s$ of the calculation.  The parameters of
  this calculation are summarized in Table \ref{tab:fiducial}.}
\label{fig:deltaE}
\end{figure} 

The first calculations we describe are summarized in Table
\ref{tab:fiducial}.  We consider an initial delta function of
particles interacting with Alfv\'enic turbulence having $\delta v
\simeq 0.5 c_s$ in the presence of a mean magnetic field with $\beta =
1$.\footnote{Weakly compressible turbulence can nonlinearly generate
  compressible fluctuations (fast waves in our case; e.g.,
  \citealt{cho2003}); this excitation is significantly weaker for
  smaller $\delta v/c_s$.  To assess whether this could be
  energetically important in our case, we carried out simulations with
  a range of $\delta v/c_s \simeq 0.1-0.5$ and found no significant
  differences relative to the results presented here, other than the
  expected scaling of the total particle heating and diffusion rates
  $\propto \delta v^2$.  This rules out excitation of fast waves as an
  important source of heating in our simulations.}  This interaction
leads to both a stochastic diffusion in velocity space and a change in
the mean energy of the particles; the magnitude of these changes
depends on the cyclotron frequency $\Omega_0$ of the particles and
their initial velocity.  Figure \ref{fig:dispersion} shows the
standard deviation of $\ug$ and $\upa$, for $\uo = c_s$ and $\mu_0 =
c_s^2/B_0$ ($u_{\perp,0} \simeq c_s$), as a function of $\Omega_0$, at
the end of the integration (after $1.0 L/c_s$). Figure
\ref{fig:deltaE} shows the parallel and perpendicular heating rates
for the same calculations $\left(\dot{E}_{\parallel,\perp} = \frac{d
    \,}{dt}\sum_{particles} \frac{1}{2} m
  u_{\parallel,\perp}^2\right)$. To make the heating rates easier to
interpret, we normalize them by the average energy dissipation rate in
the decaying turbulence.  Because our calculations are for test
particles, the heating rate can be scaled to any $\rho_{test}/\rho$,
i.e., to any test particle density.

The dispersion (Fig. \ref{fig:dispersion}) and heating
(Fig. \ref{fig:deltaE}) produced by the turbulence depend on the
gyrofrequency of the particles.  Particles with low gyrofrequency
$\Omega_0 \lesssim 300 \, c_s/L$ are the most strongly heated, with
most of the heating being in the perpendicular direction (i.e., $\ug$
increases rather than $u_\parallel$).  The heating also depends on the
gyrofrequency of the particle.  By contrast, particles with high
gyrofrequency $\Omega_0 \gtrsim 300 \, c_s/L$ have a significantly
lower heating rate that is independent of $\Omega_0$; in addition,
most of the dispersion/heating is in the direction of the magnetic
field.  We distinguish between the two classes of particles in Figures
\ref{fig:dispersion} \& \ref{fig:deltaE} by characterizing the low
$\Omega_0$ particles as ``cyclotron-resonant'' and the high $\Omega_0$
particles as ``Landau-resonant'' (the reasons for these particular
identifications will become clear shortly).

The significant heating of low $\Omega_0$ particles in Figures
\ref{fig:dispersion} \& \ref{fig:deltaE} is consistent with the
analytic expectations for cyclotron resonance (\S \ref{sec:th}).  To
confirm this, we note that the range of Alfv\'en frequencies present
in the simulation is $\sim 4-350 \, c_s/L$.  This is comparable to the
range of $\Omega_0$ in Figures \ref{fig:dispersion} \&
\ref{fig:deltaE} over which there is strong heating and diffusion.
The heating is also primarily perpendicular heating, as expected
analytically.  The negative parallel heating for the
cyclotron-resonant particles in Figure \ref{fig:deltaE} is a
consequence of pitch-angle scattering, as we demonstrate in more
detail below.

Figures \ref{fig:dispersion} \& \ref{fig:deltaE} show that particles
with high gyrofrequency $(\Omega_0 \gtrsim 10^3\,c_{s}/L)$ are
significantly less affected by the turbulence than the low $\Omega_0$
cyclotron-resonant particles.  This is particularly true for the
perpendicular velocity $\ug$: $\sigma (\ug) \ll \sigma(u_\parallel)$
at high $\Omega_0$.  We interpret this low dispersion as a nonlinear
analogue of the absence of perpendicular heating in linear theory (\S
\ref{sec:th}), which is ultimately due to the adiabatic invariance of
the magnetic moment.  In fact, $\sigma(\ug)$ at high $\Omega_0$ in
Figure \ref{fig:dispersion} is only modestly larger than the {\it
  initial} dispersion in $\ug$; because we initialize the particles
with a fixed $\mu$, there is an initial dispersion in $\ug$ due to
differences in $B$ in the turbulent plasma.

\begin{figure} 
\centering
\includegraphics[scale=0.45]{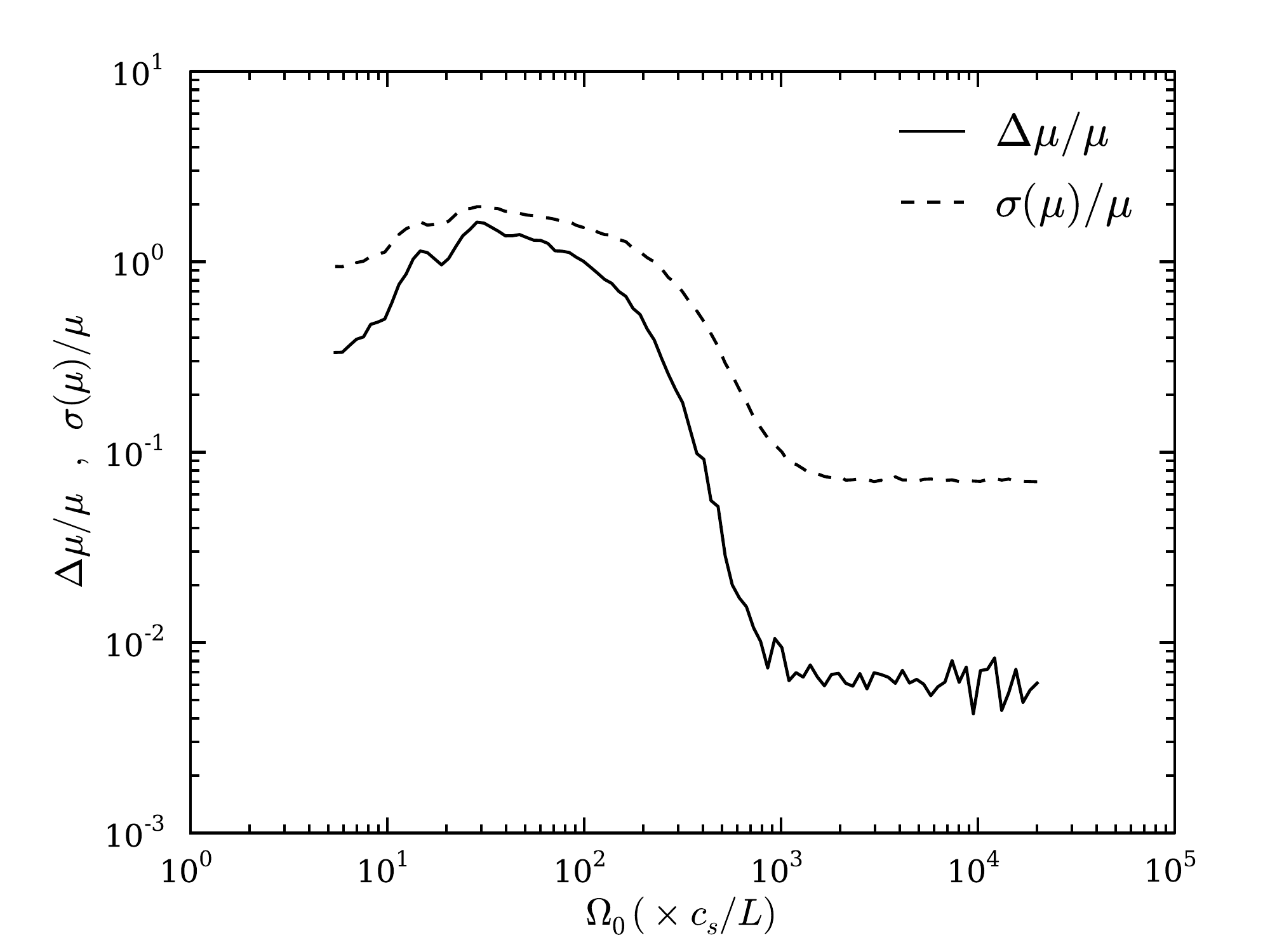}
\caption[Dispersion of $\mu$]{Standard deviation of the magnetic
  moment $\mu$ at the end of the simulation (dashed line) and the
  change in its mean (solid line), as a function of the particles'
  gyrofrequency $\Omega_0$; the particles initially have delta
  functions in velocity with $\uo = 1.0 \, c_s$ and $\mu_0 = 1.0 \,
  c_s^2/B_0$ ($\ugo \simeq 1.0 \, c_s$).  The large changes in $\mu$
  at low $\Omega_0$ are due to cyclotron resonance.  Linear theory
  predicts that $\mu$ is conserved for high $\Omega_0$, which is not
  fully consistent with the numerical results; see \S
  \ref{sec:fiducial} for an interpretation.  The parameters of this
  simulation are summarized in Table \ref{tab:fiducial}.}
\label{fig:mu}
\end{figure}

To assess the changes in $\mu$ explicitly, Figure \ref{fig:mu} shows
the standard deviation of $\mu$ and the change in the mean of $\mu$
for the same calculations as in Figures \ref{fig:dispersion} \&
\ref{fig:deltaE}.  At the end of the calculation, after 1.0 $L/c_s$,
the mean $\mu$ of the Landau-resonant (large $\Omega_0$) particles has
changed by $\lesssim 1 \%$; the dispersion introduced by the
interaction with the turbulence is somewhat larger $\sim 10 \%$.
These small changes are qualitatively consistent with the prediction
of linear theory that high gyrofrequency particles evolve
adiabatically and thus conserve their magnetic moment.  Nonetheless,
the small changes in $\mu$ in Figure \ref{fig:mu} appear to be real
and represent a quantitative difference relative to linear theory.  We
have carried out a number of tests to confirm that the changes in
$\mu$ shown in Figure \ref{fig:mu} are not numerical; e.g., the
results are independent of the timestep in the particle integrator and
of the Courant number in the integration of the turbulence.  In
addition, $\sigma (\mu) \propto \delta v$, consistent with how the
parallel diffusion depends on the amplitude of the turbulence; this
argues against violation of $\mu$ conservation due to finite amplitude
waves, as has been proposed in other contexts (e.g.,
\citealt{johnson2001}).  One possible explanation for the changes in
$\ug$ is that the delta function resonances in linear theory
(eq. [\ref{eq:res}]) are significantly broadened in the nonlinear
turbulence (as we show in more detail shortly).  Recall from equation
\eqref{eq:perp} that the change in $\ug$ due to interaction with a
single Fourier component of the turbulence is $d \ug/dt \propto
[\omega - k_\parallel u_\parallel]$.  In linear theory this vanishes
because non-zero energy exchange with a fluctuation requires $\omega =
k_\parallel u_\parallel$; this restriction is relaxed in strong
Alfv\'enic turbulence (e.g., \citealt{chandran2000}) so that small
changes in $\ug$ may be possible.

\begin{figure} 
\centering
\includegraphics[scale=0.45]{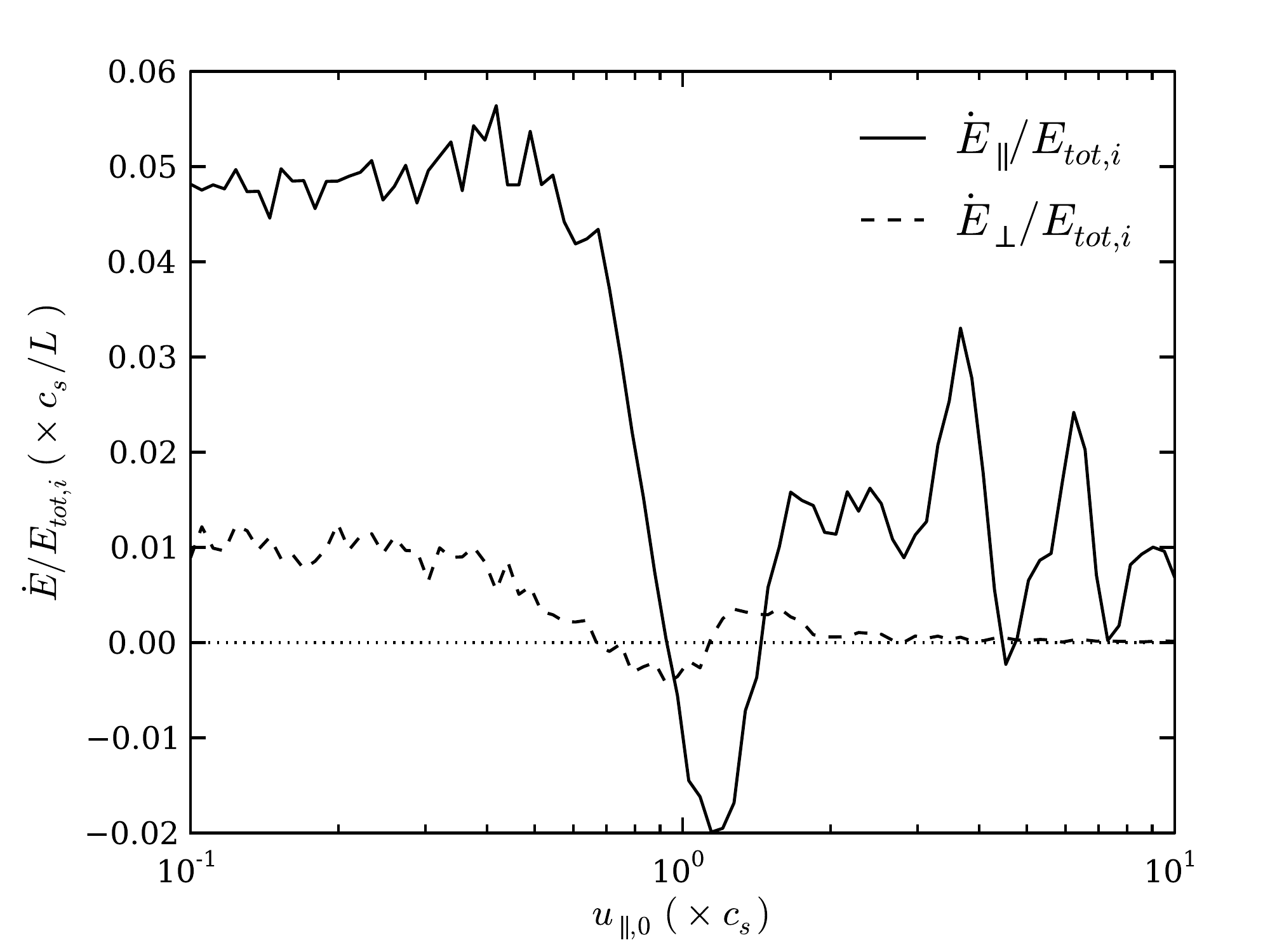}
\caption{Heating rates in the parallel (solid line) and perpendicular
  (dashed line) directions as a function of the initial parallel
  velocity $\uo$, for Landau-resonant particles $(\Omega_0 = 4.0
  \times 10^4 \, c_s/L)$. Although $\uo$ varies, all particles
  initially have the same magnetic moment, $\mu_{0} = 1.0 \,
  c_s^2/B_0$ (and thus $\ugo \simeq 1.0 \, c_s$).  Linear theory
  predicts that only particles with $\uo = \omega/k_\parallel \simeq
  0.8 c_s$ (the Landau resonance) should exchange energy with the
  waves.  By contrast, particles with all $\uo$, and particularly those
  with $\uo \lesssim c_s$, interact strongly with the turbulence. The
  heating rates are normalized by the total initial energy of the
  particle $E_{tot,i} = E_{\perp,i} + E_{\parallel,i}$, and are
  averaged over the whole integration.  The properties of this
  simulation are summarized in Table \ref{tab:fiducial}.}
\label{fig:Landau}
\end{figure} 

\begin{figure*}
\centering
\includegraphics[scale=0.44]{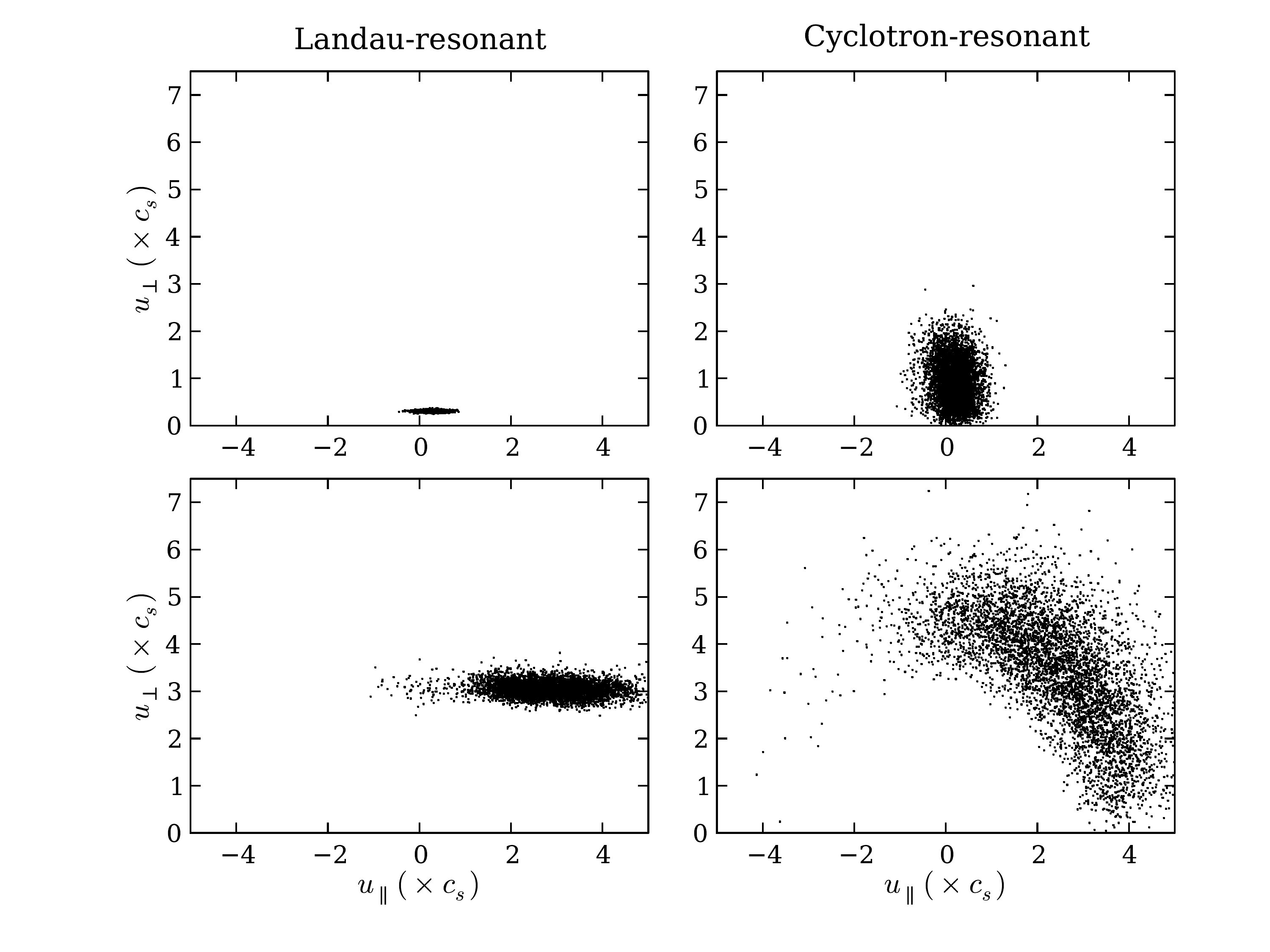}
\caption[Scatter plots in velocity space]{Scatter plots in velocity
  space ($\upa$, $\ug$) for Landau-resonant particles ($\Omega_0 = 2.0
  \times 10^4 \, c_s/L$; {\it left panels}) and cyclotron-resonant
  particles ($\Omega_0 =60 \, c_s/L$; {\it right panels}) at the end
  of the simulation.  The initial distribution function is a delta
  function.  Landau-resonant particles primarily diffuse in $\upa$
  while cyclotron resonant particles undergo pitch angle scattering
  which tends to isotropize the distribution function.  In the upper
  panels, the particles initially have low velocities, $\ugo = \uo =
  0.3 \,c_s$. In the lower panels, the particles initially have high
  velocities, $\ugo = \uo = 3.0 \,c_s$. The remaining parameters are
  summarized in Table \ref{tab:fiducial}.}
\label{fig:scatter}
\end{figure*}

As described in \S \ref{sec:th}, analytic theory predicts that high
$\Omega_0$ particles interact with anisotropic Alfv\'enic turbulence
primarily via $\mu \nabla_\parallel B$ acceleration by the slow
magnetosonic modes.  Anisotropic slow waves have the dispersion
relation $\omega \simeq c_s v_A |k_\parallel|(c_s^2 + v_A^2)^{-1/2}$
(e.g., \citealt{lg2001}), which reduces to $\omega \simeq 0.8
|k_\parallel| c_s$ for $\beta = 1$.  Thus for our fiducial
calculation, the condition for high $\Omega_0$ Landau resonance ($n =
0$ in eq. [\ref{eq:res}]) becomes: \be \upa \simeq \pm 0.8 c_{s}. \ee
Note that the resonant condition does not depend on $k$, but only on
$\upa$: the resonant particles simultaneously interact with all modes,
but only a single parallel velocity is picked out in linear theory.
Figure \ref{fig:Landau} shows the parallel and perpendicular heating
as a function of the initial parallel velocity $\uo$ for our fiducial
calculation, taking a fixed $\mu_0 = 1.0 \, c^2_s/B_0$ and a sufficiently
high $\Omega_0 = 4.0 \times 10^4 \, c_s/L$ that there is no cyclotron
resonance.  Contrary to linear theory, particles with a wide range of
velocities $\sim 0.1-10 \, c_s$ receive significant energy; in
particular, for $\uo \lesssim 0.8 c_s$ (the resonant velocity in
linear theory), the heating is relatively independent of
$\uo$.\footnote{Note that the results for Figures \ref{fig:dispersion}
  and \ref{fig:deltaE} were for $u_{\parallel,0} = c_s$, for which the
  heating is particularly small.} The wide range of velocities at
which the particles couple to the turbulence is qualitatively
consistent with the idea that the linear resonances
(eq. [\ref{eq:res}]) are highly broadened in Alfv\'enic turbulence
because the nonlinear decorrelation time is comparable to or shorter
than the linear period of the Alfv\'en and slow waves (e.g.,
\citealt{chandran2000}).  We defer a more detailed test of resonance
broadening models to future work.  Although the linear resonance no
longer manifests itself as a delta function, Figure \ref{fig:Landau}
shows that at nearly all velocities, the heating of high $\Omega_0$
particles is primarily parallel to the local magnetic field,
consistent with linear theory predictions.

Figure \ref{fig:scatter} shows the positions of the individual
particles in velocity space at the end of the fiducial simulation, for
both high $\Omega_0$ Landau-resonant particles (left column) and low
$\Omega_0$ cyclotron-resonant particles (right column).  The top row
is for low velocity particles with $\uo \simeq \ugo \simeq 0.3 \, c_s$
while the bottom row is for high velocity particles with $\uo \simeq
\ugo \simeq 3 \, c_s$.  For Landau-resonant particles, the dispersion
occurs almost exclusively in the parallel direction, as we have
previously seen.  For the cyclotron-resonant particles, the diffusion
in velocity space depends somewhat on the velocity of the
particles. For $\uo \lesssim v_{A}$ (top right), the particles are
preferentially dispersed in the perpendicular direction due to the
cyclotron resonance. However for $\uo > v_{A}$ (bottom right), the
velocity diffusion is dominated by \emph{pitch angle scattering}: the
velocity distribution is isotropized more rapidly than the total
energy changes.  This difference in the results of cyclotron resonance
for sub and super-Alfv\'enic particles is a well-known result of
quasilinear theory (e.g., \citealt{miller1995}).

\subsection{Resistive Simulations}\label{sec:resistivity}

\begin{figure} 
\centering
\includegraphics[scale=0.45]{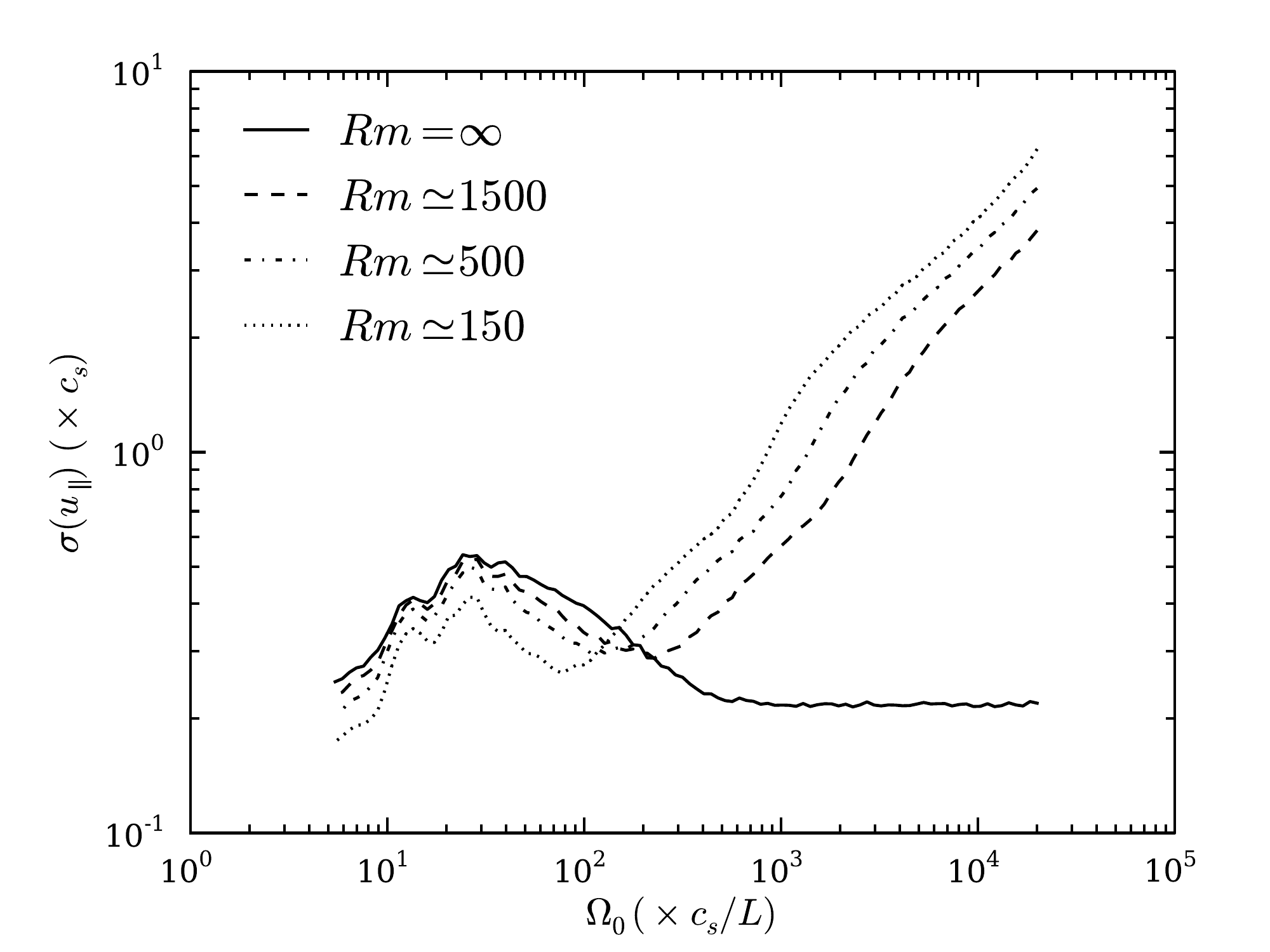}
\includegraphics[scale=0.45]{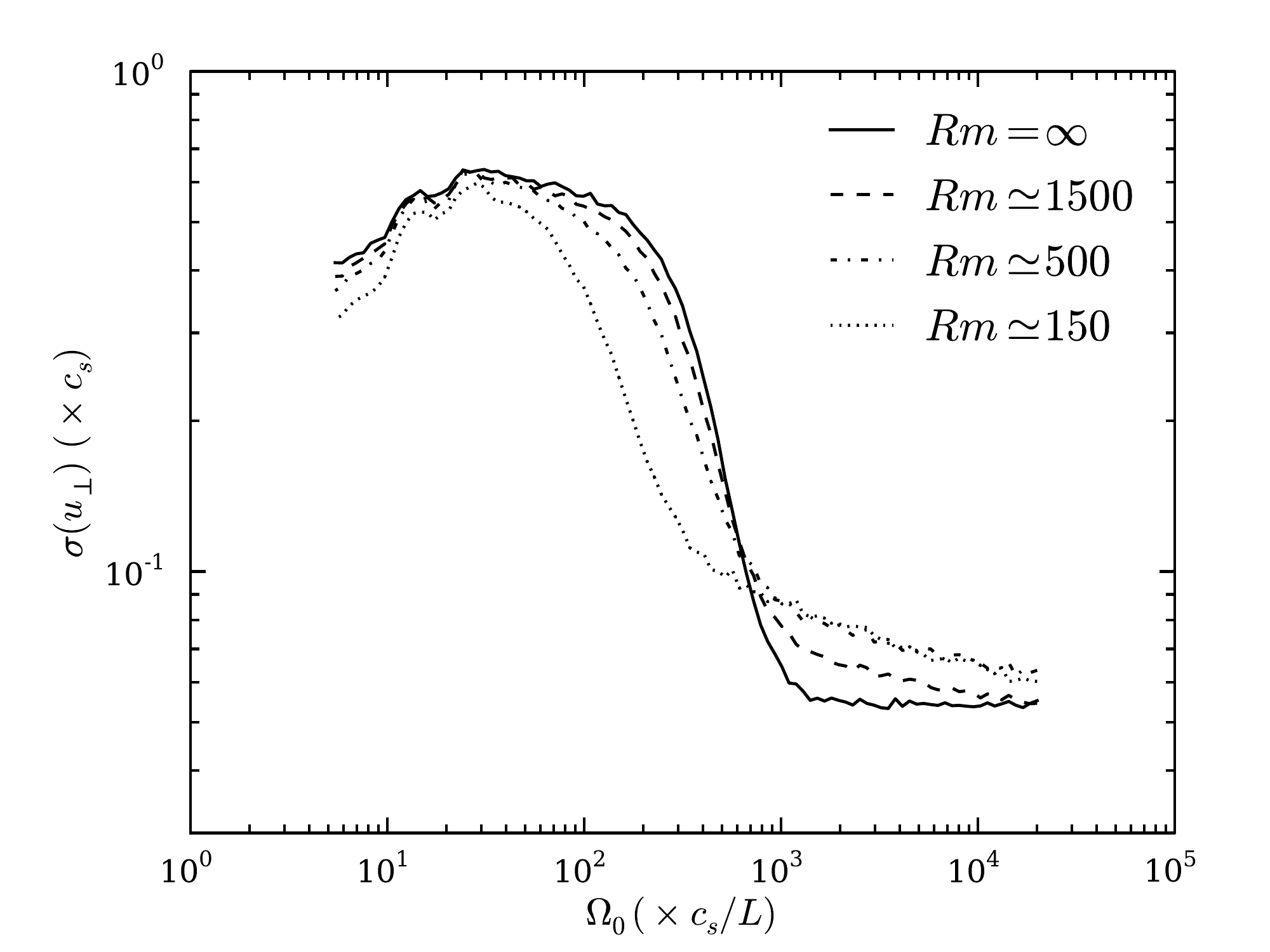}
\caption[Dispersion for different resistivities]{Standard deviation of
  the parallel $\upa$ (top panel) and perpendicular $\ug$ (bottom
  panel) velocities at the end of the simulation, as a function of the
  particles' gyrofrequency $\Omega_0$.  The curves are plotted for
  different resistivities $\eta$, which are converted to magnetic
  Reynolds numbers, as computed \emph{on the outer scale of the
    turbulence}. The solid line is for $\eta = 0$ (ideal MHD), while
  the dashed, dot-dashed and dotted lines have increasing resistivity.
  The presence of $\eta \ne 0$ produces an $E_\parallel$ that leads to
  significant parallel heating and diffusion at high $\Omega_0$, but
  the properties of the low $\Omega_0$ cyclotron-resonant particles do
  not depend significantly on $\eta$.  The particles initially have
  delta functions in velocity with $\uo = 1.0 \, c_s$ and $\mu_0 = 1.0
  \, c_s^2/B_0$ ($\ugo \simeq 1.0 \, c_s$).  Aside from $\eta$, the
  parameters of this simulation are the same as in Table
  \ref{tab:fiducial}. }
\label{fig:resistive}
\end{figure}

In the ideal MHD simulations described in the previous sub-section,
$E_\parallel = 0$ because of our constrained transport algorithm; this
is also preserved to machine accuracy by our interpolation methods (\S
\ref{sec:numerics}).  Numerical reconnection is nonetheless present on
small scales. To understand the effects of reconnection and
small-scale current sheets on our results in a more controlled manner,
we have carried out a number of simulations with finite resistivity
$\eta$.\footnote{We keep the energy injection rate $\dot{E}$ constant,
  which leads to a slightly lower Mach number for larger $\eta$, due to
  the higher dissipation.}  It is important to stress that the physics
of reconnection is not adequately represented by a spatially and
temporally constant value of $\eta$.  Thus our calculations with
finite $\eta$ should not be interpreted as physical.  Rather, they
allow us to isolate the potential importance of reconnection for some
of our results and assess whether the interpretations given in the
previous sub-section are correct.

We parameterize our chosen values of $\eta$ using the magnetic
Reynolds number of the turbulence on the outer scale: $Rm \equiv L
\delta v/[\eta c/4\pi]$.  We chose values of $\eta$ so that $Rm \sim
1$ \emph{on the grid scale}, i.e., so that small-scale current sheets
are marginally resolved; this implies \be \frac{ \Delta x \, \delta
  v_{k_{max}} }{ (\eta c/4\pi) }\sim 1 \ee where $\Delta x$ is the
size of a cell and $k_{max} \simeq \pi/\Delta x$. We estimate $\delta
v_{k_{max}}$ assuming a Kolmogorov cascade $\delta v_{k} \approx
\delta v (k L)^{-1/3}$.  For our fiducial resolution of $256^3$, $Rm
\sim 1$ on the grid scale thus corresponds to $Rm \sim 1500$ at the
outer scale of the turbulence.

Figure \ref{fig:resistive} compares the velocity dispersion in the
parallel and perpendicular directions at the end of simulations with
different values of $\eta$, for particles with $\uo = 1.0 \, c_s$ and
$\mu_0 = 1.0 \, c_s^2/B$.  The most striking result is the very strong
parallel dispersion for particles with high gyrofrequency. This is a
consequence of the non-zero parallel electric field in resistive MHD:
$E_{\parallel} = \frac{\eta}{4\pi} (\bs{\nabla}\times
\bs{B})_{\parallel}$. As a result, particles are freely accelerated in
the parallel direction by the $qE_{\parallel}$ force (see eq.
[\ref{eq:para}]). The parallel acceleration thus depends strongly on
the particle's charge-to-mass ratio $q/m$ -- or equivalently on its
$\Omega_0$.  This dependence is clearly visible in Figure
\ref{fig:resistive}.\footnote{We found similar results in \emph{ideal}
  MHD, when we did not explicitly constrain $E_{\parallel}$ to be zero
  when interpolating from the MHD grid to the particle's position (see
  \S \ref{sec:interpolation}).}

The perpendicular velocity $\ug$ is much less affected by resistivity,
since the additional perpendicular electric field $\bs{E}_{\perp} =
\frac{\eta}{4\pi} (\bs{\nabla}\times \bs{B})_{\perp}$ averages to zero
over a gyration. The strongest effect evident in Figure
\ref{fig:resistive} is that, for $\Omega_0 \sim 100 \,c_s/L$, the
perpendicular diffusion becomes weaker as the resistivity
increases. The most probable explanation is that resistive dissipation
preferentially damps the high $k$ modes, which are precisely the modes
that resonate with particles having $\Omega_0 \sim 100 \,c_s/L$.

Overall, our simulations with an explicit resistivity demonstrate
that, in resistive MHD, current sheets lead to parallel heating by the
non-zero $E_\parallel$.  This is in addition to the parallel heating
by $\mu \nabla_\parallel B$ forces and the perpendicular heating and
pitch-angle scattering by cyclotron-resonant waves highlighted in the
previous sub-section.

\subsection{Variations in $\beta$}\label{sec:beta}

\begin{figure} 
\centering
\includegraphics[scale=0.45]{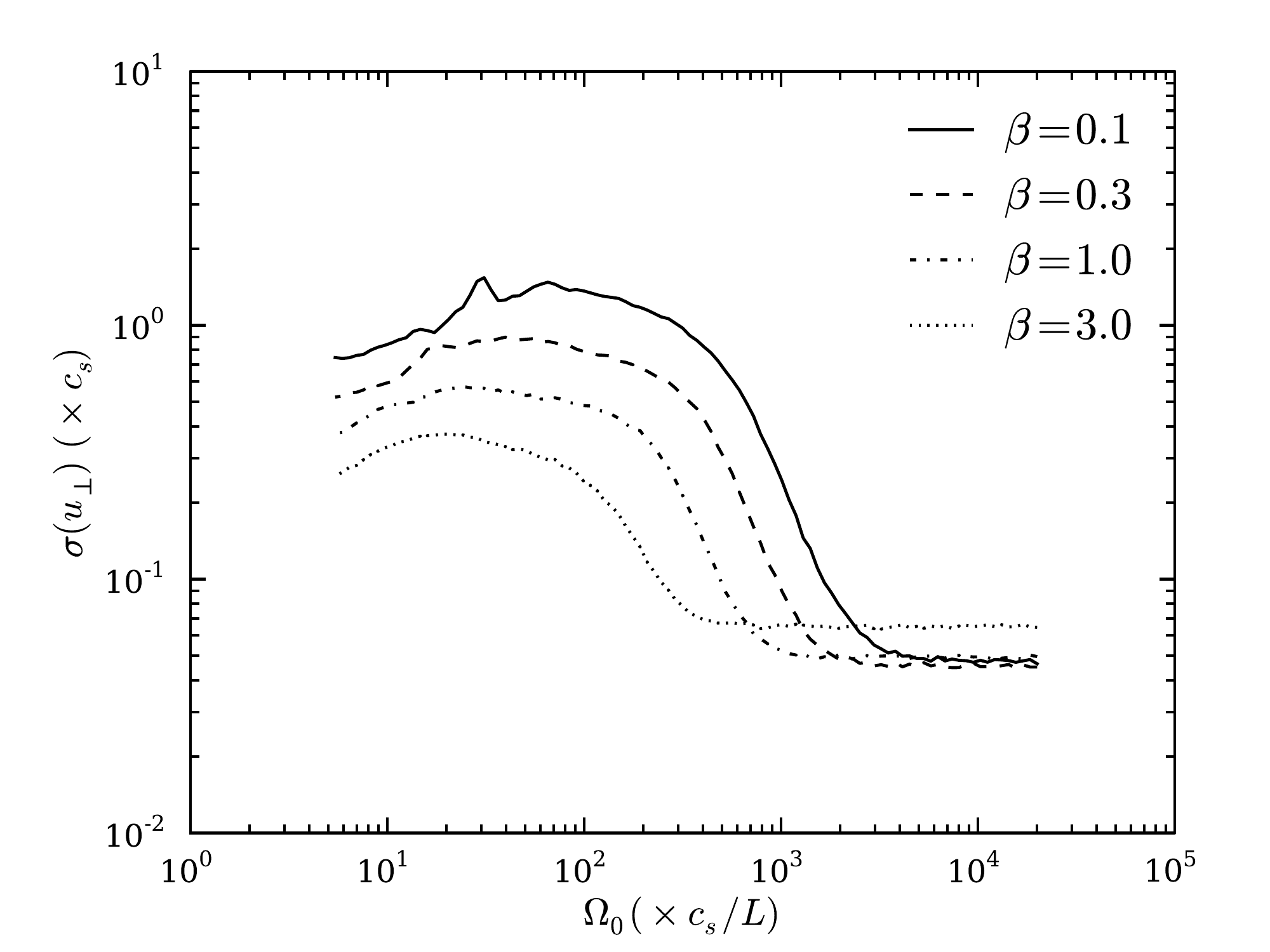}
\caption[Dispersion for different $\beta$'s]{Standard deviation of the
  perpendicular velocity $\ug$ at the end of the simulation as a
  function of the particles' gyrofrequency $\Omega_0$, for $\uo = 0$
  (to minimize Doppler shift) and $\ugo \simeq 1.0 \, c_s$.  The
  different curves correspond to different values of $\beta$.  Linear
  theory predicts that cyclotron resonance with Alfv\'en waves
  requires $|k_\parallel| v_A \simeq \Omega_0$, consistent with the
  increasing heating/dispersion at high $\Omega_0$ for decreasing
  $\beta$ (i.e., increasing $v_A$). $\dot E$ is adjusted so that all
  calculations have approximately the same Alfv\'enic Mach number.}
\label{fig:beta}
\end{figure}

We carried out a number of calculations at different $\beta$ to assess
whether the particle heating physics identified in \S
\ref{sec:fiducial} depends significantly on $\beta$.  Because we are
focusing on weakly compressible Alfv\'enic turbulence, we kept the
Alfv\'enic Mach number $\delta v/v_{A}$ at the saturation of the
turbulence constant when varying $\beta$ (by appropriately varying
$\dot{E}$). There were no significant changes in any of our results as
a function of $\beta$.  To illustrate one example, Figure
\ref{fig:beta} shows how the standard deviation of $\ug$ at the end of
the simulation depends on the background field.  In this calculation,
the particles initially have $\mu_0 = 1.0 \, c^2_s/B_0$ and $\uo = 0$,
the latter to avoid any Doppler shifts in the cyclotron resonance.

For cyclotron resonant particles, the decreasing magnitude of
$\sigma(\ug)$ for higher $\beta$ in Figure \ref{fig:beta} is due to
the fact that, to keep $\delta v/v_{A}$ constant, the energy in the
turbulent fluctuations is lower for high $\beta$.  The cyclotron
resonance clearly shifts towards higher $\Omega_0$ as $\beta$
decreases.  This can be understand from equation \eqref{eq:resonance}
and the assumption of Alfv\'enic turbulence: resonance requires
$|k_\parallel| v_A \sim \Omega_0$ and thus the resonant particles
should have $\Omega_0 \propto \beta^{-1/2}$.  This scaling is reasonably
consistent with the shift in Figure \ref{fig:beta}.

\subsection{The Effects of Numerical Resolution}\label{sec:resolution}

\begin{figure} 
\centering
\includegraphics[scale=0.45]{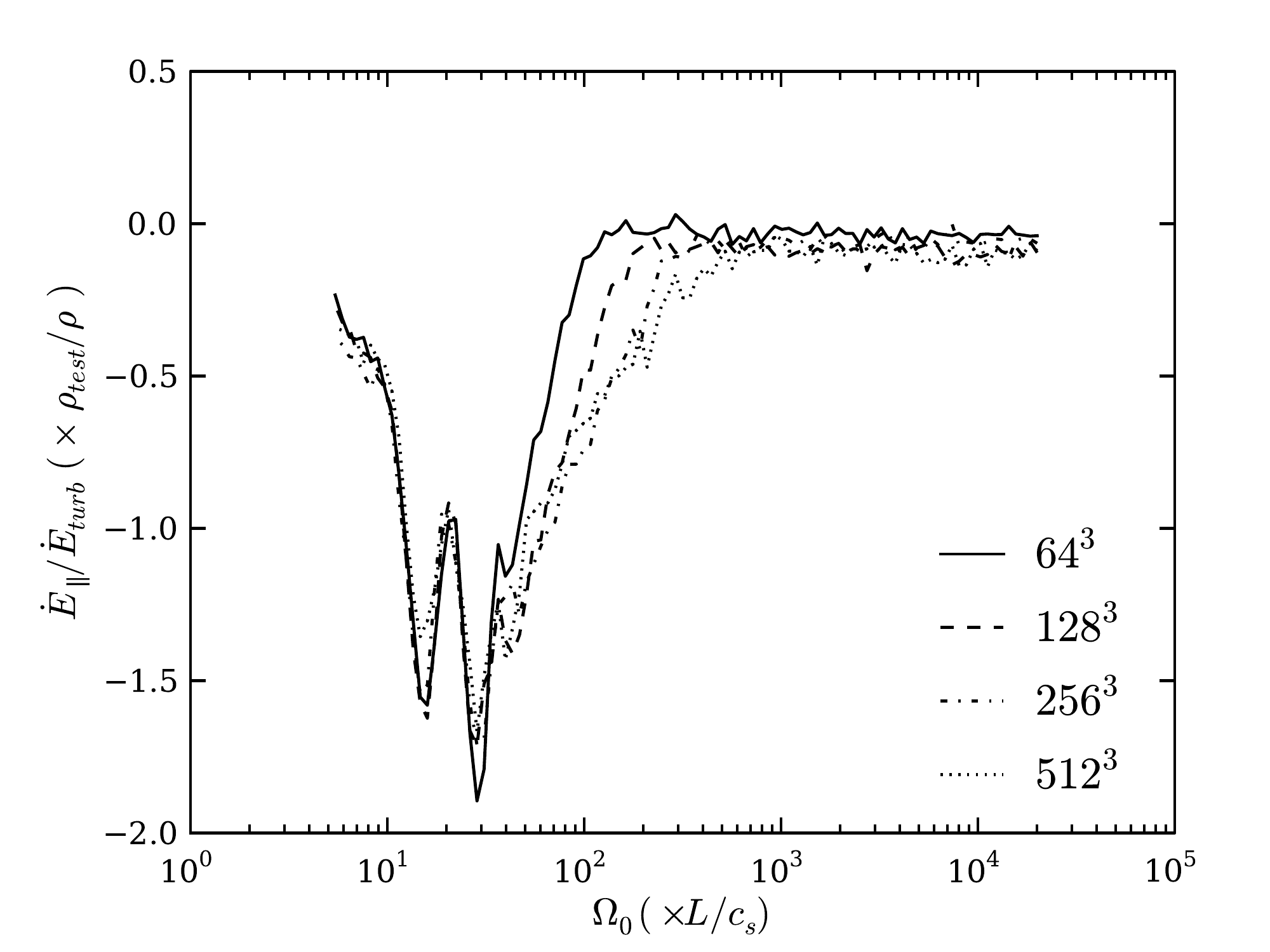}
\includegraphics[scale=0.45]{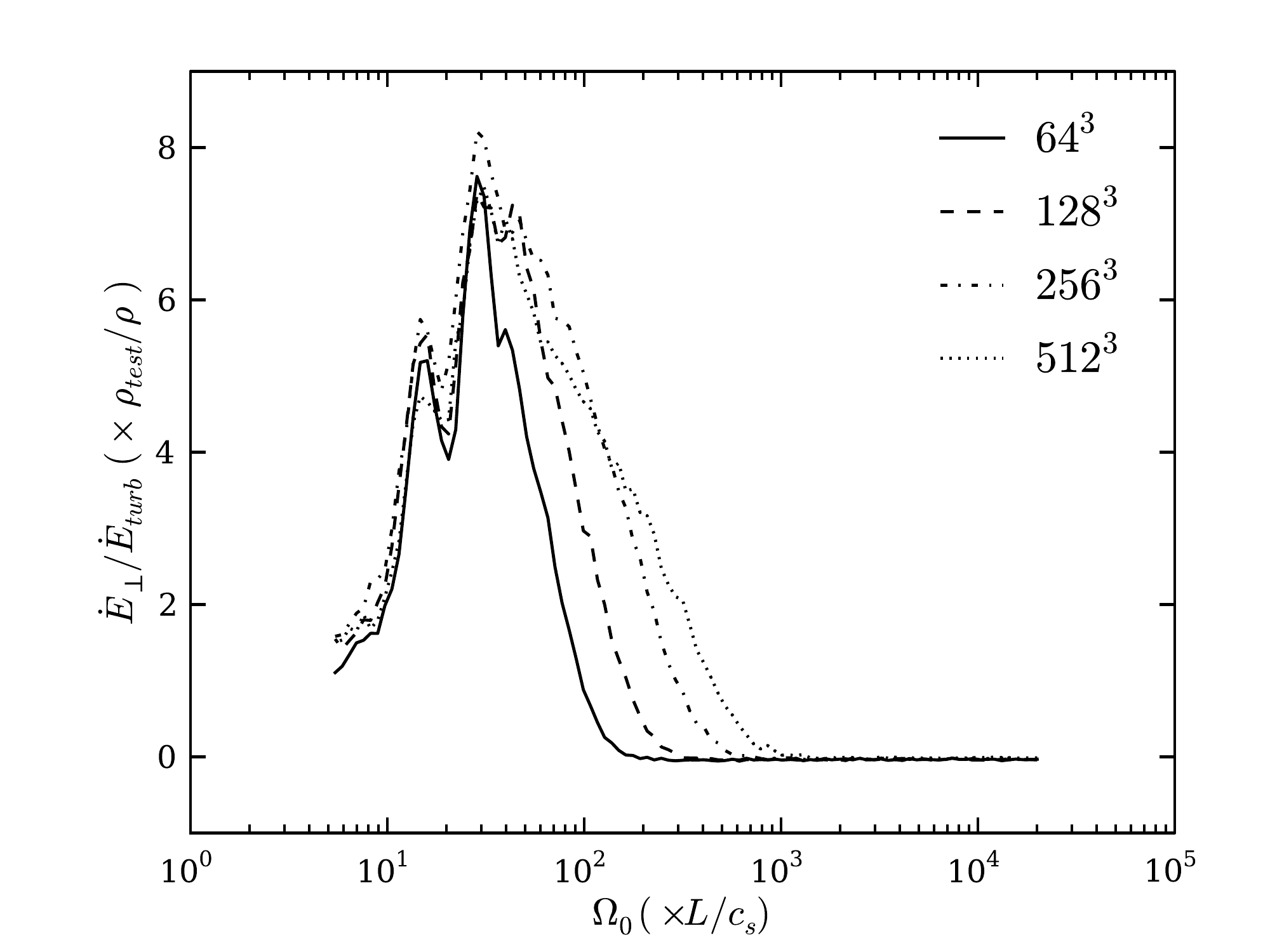}
\caption[Resonance of $\ug$ for different resolutions]{Parallel (upper
  panel) and perpendicular (lower panel) heating rates, as a function
  of the particles' gyrofrequency $\Omega_0$, for $\uo = 1.0 \, c_s$
  and $\ugo \simeq 1.0 \, c_s$.  The different curves correspond to
  different resolutions. With increasing resolution, higher
  frequencies are present on the computational domain, which increases
  the range in $\Omega_0$ over which there is perpendicular heating
  due to the cyclotron resonance.  By contrast, the parallel heating
  at high $\Omega_0$ is relatively independent of resolution.  The
  heating rates are averaged over the whole integration (i.e., for
  $1.0\, L/c_s$), and are normalized by $\dot{E}_{turb}$, the energy
  dissipation rate in the turbulence.  Apart from resolution, the
  properties of these calculations are summarized in Table
  \ref{tab:fiducial}.}
\label{fig:resolution}
\end{figure}

Finite numerical resolution limits the conclusions we can draw for
several reasons, even given our restriction to scales where MHD is
valid.  In particular, higher resolution changes the properties of the
turbulence, since we can resolve higher $k$ modes.  For the same
reason higher resolution increases the range of linear and nonlinear
frequencies found in the turbulence, and thus the range of particles
that can be cyclotron resonant.

Figure \ref{fig:resolution} shows the parallel and perpendicular
heating rates for different resolutions. Going to higher resolution
does not significantly affect the parallel heating rate at high
$\Omega_0$.  This is reasonably consistent with linear theory:
\citet{barnes1966} showed that the linear collisionless damping of the
slow magnetosonic mode is independent of the magnitude of ${\bf k}$
(although it depends on its direction).  Thus increasing the range of
${\bf k}$ contained in the domain should not significantly change the
heating rate, since all scales contribute significantly.  By contrast,
Figure \ref{fig:resolution} shows that the range of gyrofrequencies
$\Omega_0$ that can be cyclotron resonant increases with resolution:
higher frequency particles are cyclotron-resonant at higher resolution
because the range of resolved ${\bf k}$, and thus the range of
resolved frequencies, increases at higher resolution.

%\vspace{0.66cm}

\begin{figure} 
\centering
\includegraphics[scale=0.45]{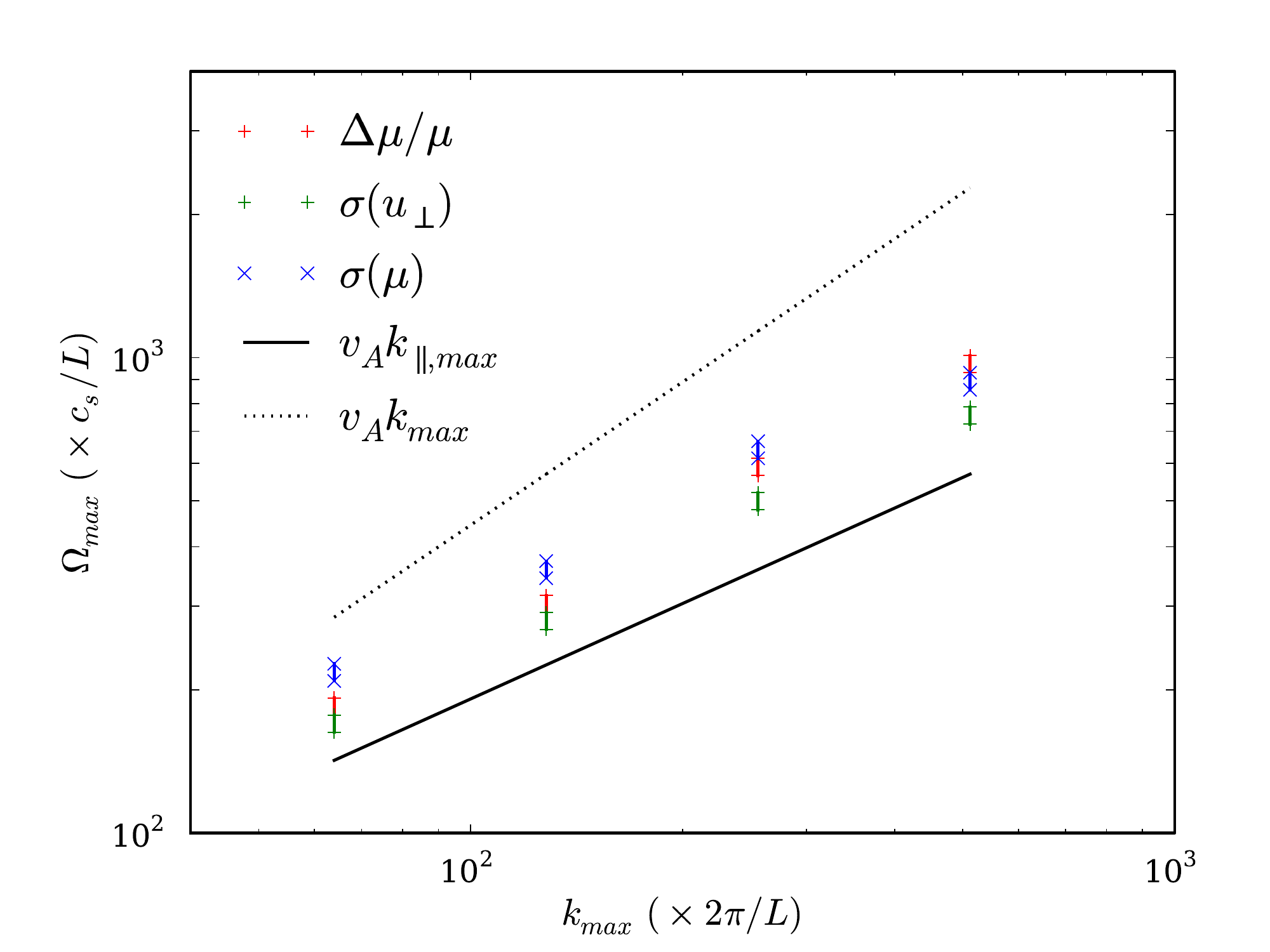}
\caption[kmax scaling]{Maximum gyrofrequency $\Omega_{max}$ for
  cyclotron-resonant particles as a function of resolution (colored
  symbols); the resolution is quantified by the maximum wavenumber
  resolved in the simulation $k_{max} \equiv \pi/\Delta x$.  Also
  shown is the maximum frequency of Alfv\'en waves as a function of
  resolution in an isotropic cascade (dotted line) and for anisotropic
  Alfv\'enic turbulence (solid line; $k_{\parallel max} \simeq
  k_{max}^{2/3} k_{min}^{1/3}$).  The numerical results favor the
  anisotropic turbulence models.  $\Omega_{max}$ is the value of the
  particle gyrofrequency $\Omega_0$ at which $\Delta \mu/\mu$,
  $\sigma(\ug)$, or $\sigma(\mu)$ is larger than its $\Omega_0
  \rightarrow \infty$ value by a factor $\sim 2$ (see \S
  \ref{sec:resolution} for details).}
\label{fig:max}
\end{figure}

Quantifying the range of $\Omega_0$ that are cyclotron-resonant at a
given resolution is critical for understanding the implications of our
results for solar and astrophysical problems (\S
\ref{sec:implication}).  Towards this end we define a maximum
cyclotron frequency $\Omega_{max}$: this is the value of $\Omega_0$ at
which $\Delta \mu/\mu$, $\sigma(\ug)$, or $\sigma(\mu)$ is larger than
its $\Omega_0 \rightarrow \infty$ value by a factor of $2$.  The exact
definition of $\Omega_{max}$ is somewhat arbitrary, but our definition
captures the key result seen in many of the plots in this paper (e.g.,
Fig. \ref{fig:dispersion} \& \ref{fig:mu}): the dispersion/heating in
$\mu$ and $\ug$ are small at high $\Omega_0$ (for Landau-resonant
particles) and then increase rapidly for smaller $\Omega_0$ as
$\Omega_0$ becomes comparable to the frequency of Alfv\'en waves on
the computational domain.  By considering three different physical
quantities ($\Delta \mu/\mu$, $\sigma(\ug)$, and $\sigma(\mu)$) in
defining $\Omega_{max}$ we hope to bracket some of the uncertainty in
our estimate of $\Omega_{max}$.

Figure \ref{fig:max} shows these three different definitions of
$\Omega_{max}$ as a function of $k_{max} \equiv \pi/\Delta x$, for our
fiducial calculation with $\beta = 1$, $\uo = 1.0 \, c_s$, and $\ugo
\simeq 1.0 \, c_s$.  There is a clear tend of increasing
$\Omega_{max}$ with $k_{max}$.  Also shown in Figure \ref{fig:max} are
two theoretical predictions for the maximum Alfv\'en wave frequency as
a function of resolution.  The first (dotted line) assumes an
isotropic cascade in which case the maximum Alfv\'en frequency is
$\simeq k_{max} \, v_A$.  The second estimate (solid line) takes into
account the anisotropy of Alfv\'enic turbulence; critical balance
implies that the maximum Alfv\'en frequency is $k_{\parallel, max} v_A
\simeq k_{max}^{2/3} \, k_{min}^{1/3} \, v_A$, where in Figure
\ref{fig:max} we take $k_{min} = 2 \pi/(L/4)$ since we drive modes
with wavelengths between $L/4$ and $L$ (see \S \ref{sec:MHD}).

The estimates of $\Omega_{max}$ in Figure \ref{fig:max} are reasonably
consistent with $\Omega_{max} \simeq 2 \, k_{\parallel, max} v_A$,
which is comparable to the maximum Alfv\'en wave frequency in {\it
  anisotropic} Alfv\'enic turbulence.  Because the discrete resonances
of linear theory are ``broadened'' in strong MHD turbulence
(Fig. \ref{fig:Landau}), the maximum gyrofrequency of particles that
feel the cyclotron resonance should be somewhat {\it larger} than
maximum Alfv\'en wave frequency.  This is true for the anisotropic
estimate of $\Omega_{max}$ in Figure \ref{fig:max}, but not the
isotropic estimate.  The non-trivial point demonstrated by Figure
\ref{fig:max} is that (as predicted by linear theory) it is the
smallest {\it parallel} length-scale in the turbulent fluctuations
that determines the efficacy of cyclotron resonance, and thus
perpendicular heating.  This fact will be very important when we apply
our results to space physics and astrophysics problems in \S
\ref{sec:implication}.

\subsection{Heating of a Thermal Distribution}\label{sec:thermal}

In the previous sections we have focused on the diffusion and heating
of particles that all start with approximately the same velocity, in
order to isolate the physics of test particles interacting with strong
Alfv\'enic turbulence.  In order to obtain a more realistic estimate
of the total heating rate in a turbulent plasma, we now consider an
isotropic thermal distribution of particles ($f_{MB}$ in
eq. [\ref{eq:thermal}]).  These calculations have $\beta = 1$ and the
test particles have a thermal speed equal to that of the fluid; thus
the thermal test particles represent protons or other particles with a
similar thermal speed.

\begin{figure} 
\centering
\includegraphics[scale=0.45]{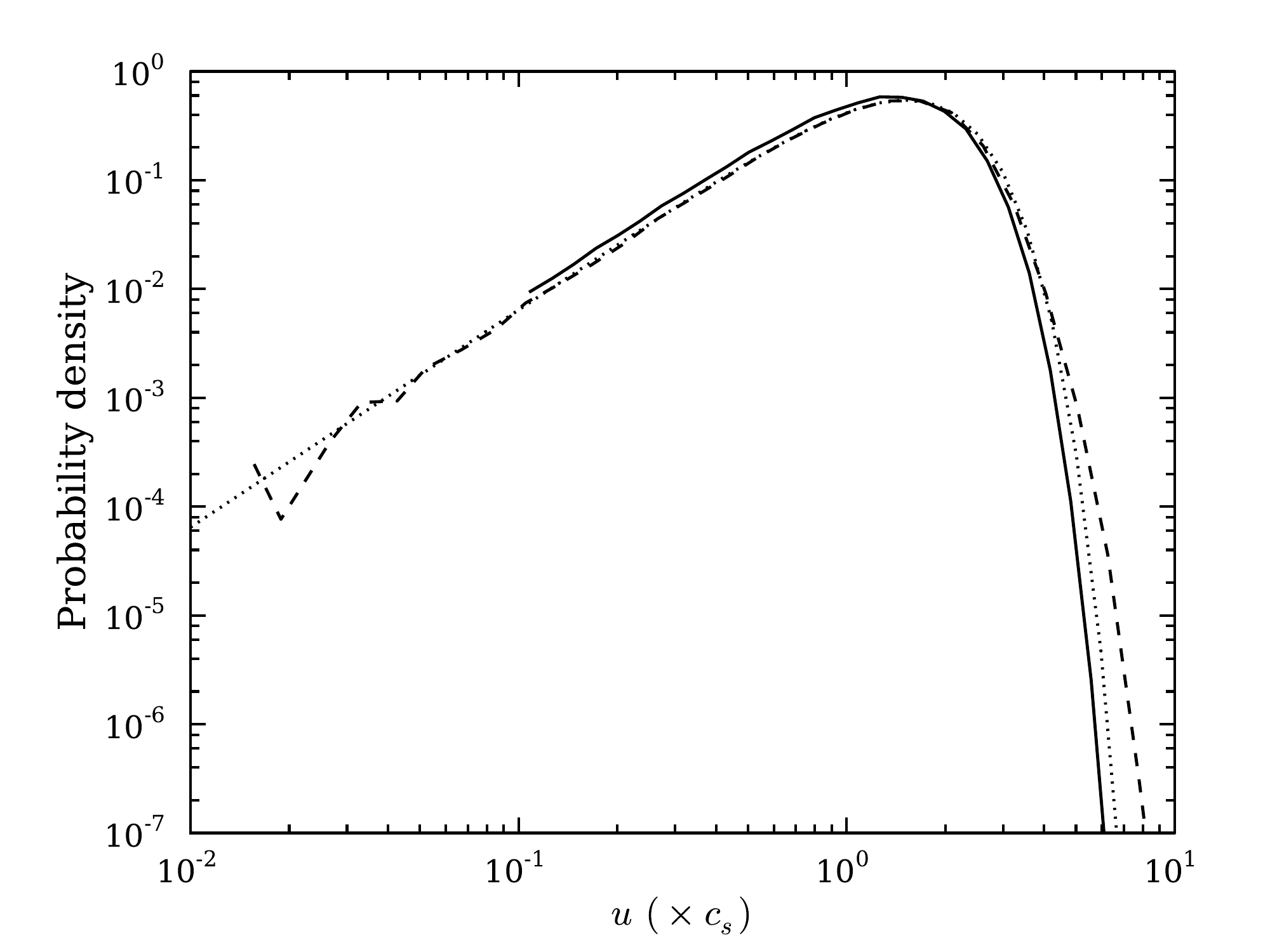}
\caption[Velocity distribution in log space]{Distribution functions
  vs. total velocity $u = \sqrt{ \ug^2 + \upa^2}$, for
  cyclotron-resonant particles with $\Omega_0 = 250 \, c_s/L$.  The
  distribution function is initially thermal with the same sound speed
  as the fluid ({\it solid line}); the test particles thus represent
  protons.  The {\it dashed line} is the distribution function at the
  end of the integration (after $1.0 L/c_s$), while the {\it dotted
    line} is a thermal distribution with the same average energy as
  the final distribution.  The cyclotron resonance leads to modest
  acceleration of high energy particles.  The properties of the
  simulation are the same as in Table \ref{tab:fiducial}, except for
  the resolution, which is $512^3$. }
\label{fig:distribution}
\end{figure}

\begin{figure} 
\centering
\includegraphics[scale=0.45]{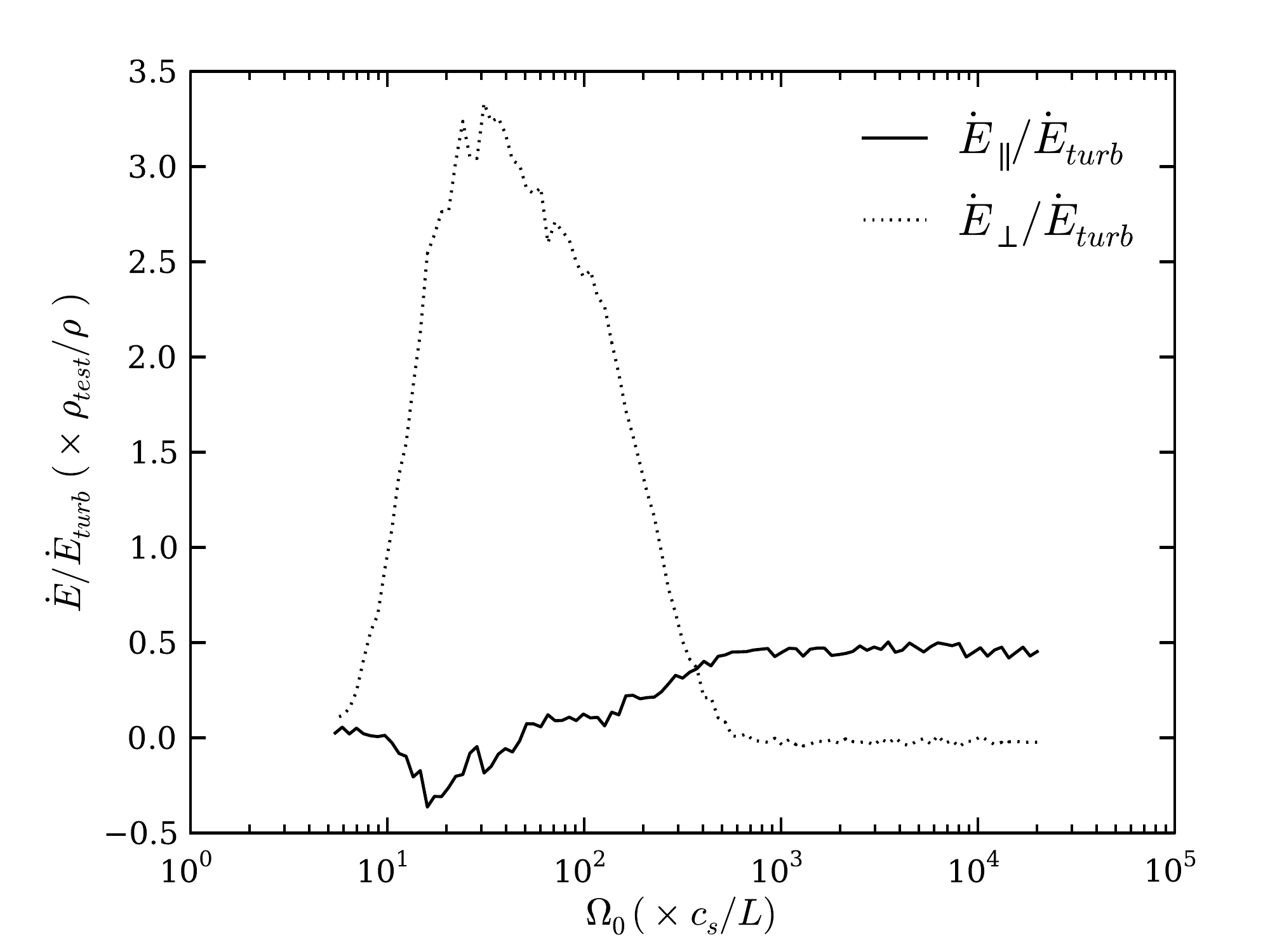}
\caption[Rescaled particle heating rate]{Parallel (solid line) and
  perpendicular (dashed line) heating rates, as a function of the
  particles' gyrofrequency $\Omega_0$.  The heating rates are averaged
  over the last half of the simulation (i.e., for $0.5\, L/c_s$), and are
  normalized by $\dot{E}_{turb}$, the energy dissipation rate in the
  turbulence.  The distribution function is initially thermal with the
  same sound speed as the fluid; the test particles thus represent
  protons.  The Landau-resonant particles with high $\Omega_0$ receive
  primarily parallel heating while the cyclotron-resonant particles
  with lower $\Omega_0$ are largely heated in the direction
  perpendicular to the local magnetic field.}
\label{fig:Edot}
\end{figure}

Figure \ref{fig:distribution} shows the evolution of the velocity
distribution as the simulation proceeds, for cyclotron-resonant
particles $(\Omega_0 = 250\,c_s/L)$. The distribution function
develops a modest non-thermal tail, although most of the energy
remains in the thermal population of particles.  Figure \ref{fig:Edot}
shows the particles' parallel and perpendicular heating rates, divided
by the average energy dissipation rate in the turbulence, as a
function of the particles' gyrofrequency $\Omega_0$. As before, these
results can be rescaled by $\rho_{test}/\rho$ to apply them to any
specific population of test particles.  For high gyrofrequency
$(\gtrsim 500 \,c_s/L )$, Figure \ref{fig:Edot} shows that the heating
is primarily parallel to the magnetic field; the total perpendicular
heating rate fluctuates around $\simeq - 0.025 \, \dot E_{turb} \simeq
- \dot E_{\parallel}/20$, sufficiently small that it is hard to read
off of the Figure.  As we have argued before, this parallel heating is
due to the stochastic $\mu \nabla B$ forces created by slow
magnetosonic waves.  Note that the magnitude of the parallel heating
rate is $\simeq 0.5 \, \dot E_{turb}$.  This is sufficiently strong
that, for the parameters chosen here ($\beta = 1$), a significant
fraction of the energy in the slow wave cascade would be dissipated on
large scales, rather than cascading to small scales.

For lower $\Omega_0$ ($10 - 300 \,c_s/L$), Figure \ref{fig:Edot} shows
that the net heating of the thermal distribution function is primarily
perpendicular to the magnetic field.  The heating is very strong,
$\simeq 3 \dot E_{turb}$ (for $\rho_{test} \simeq \rho$) and $\simeq
6$ times the heating rate in the high $\Omega_0$ limit.  This strong
perpendicular heating is, we have argued, due to cyclotron resonance.
For $\rho_{test} \simeq \rho$, the energy gained by the particles
significantly exceeds the energy available in the turbulence. This
clearly demonstrates that the damping is sufficiently strong that the
test particle approximation breaks down, if in fact the turbulent
fluctuations reach $\omega \sim \Omega_0$.  We return to this issue in
\S \ref{sec:disc}.

\section{Summary and Implications}

\label{sec:disc}

We have carried out a detailed study of the heating of test particles
by weakly compressible MHD turbulence.  Our calculations integrate the
equations of motion for up to $\sim 10^6$ particles in ``real time''
as the turbulence itself evolves.  The heating and acceleration of
particles by MHD turbulence plays a central role in theoretical models
of many heliospheric and astrophysical phenomena, including solar
flares \citep{miller1995}, cosmic-ray scattering and confinement in
the Galaxy \citep{chandran2000}, the heating and acceleration of the
solar wind \citep{cranmer2005}, and the radiation produced by some
accretion disks onto compact objects \citep{Quataert:1999}.  Because
of the many potential applications of this work, we have focused on
the basic physics of particles interacting with MHD turbulence, rather
than on any of these specific applications.  In this section we first
summarize our primary results and then briefly discuss their
implications for astrophysical systems, in particular the solar wind.

Our turbulence is driven subsonically, sub-Alfv\'enically and with a
divergence free velocity field (\S \ref{sec:MHD}).  Such weakly
compressible MHD turbulence consists of nonlinearly interacting
Alfv\'en and slow magnetosonic waves, with most of the energy
cascading to small scales perpendicular to the local magnetic field
(e.g., \citealt{GS95}). There is an extensive body of work studying
the heating of test particles using (quasi)linear theory, which
predicts that energy exchange happens at discrete resonances (e.g.,
\citealt{stix1992} and references therein); we review these
predictions in \S \ref{sec:th}.  Our calculations relax many of the
simplifying assumptions made in linear theory.  Specifically, rather
than studying the interaction between particles and a superposition of
long-lived linear waves, our particles interact with strong MHD
turbulence, i.e., with turbulence in which the timescale for nonlinear
interactions to transfer energy to smaller scales is comparable to, or
shorter than, the linear period of the waves.  We have also carried
out both ideal and resistive-MHD simulations in order to isolate the
importance of dissipation in current sheets, rather than wave-particle
resonances, for particle heating.

\subsection{Summary}

One of the striking conclusions of our work is that many -- although
not all -- of the results we find for test particle diffusion and
heating in fully developed MHD turbulence are qualitatively consistent
with the predictions of quasilinear theory (we defer a more
quantitative comparison between our numerical results and quasilinear
theory to future work).  More specifically, our primary results can be
summarized as follows:

\noindent $\bullet$ How particles are heated depends sensitively on
whether the gyrofrequency of the particle $\Omega_0$ is comparable to
the frequency of a turbulent fluctuation $\omega$ that is resolved on
the computational domain.

\noindent $\bullet$ Particles with $\Omega_0 \sim \omega$ undergo
strong {\it perpendicular} heating and {\it pitch angle scattering},
qualitatively consistent with linear theory predictions for cyclotron
resonance with the linearly polarized Alfv\'en waves present in MHD
turbulence (e.g., Fig. \ref{fig:dispersion}, \ref{fig:deltaE},
\ref{fig:scatter}, \& \ref{fig:Edot}).  In \S \ref{sec:implication} we
discuss the implications of this perpendicular heating for
measurements of proton and ion temperature anisotropies in the solar
corona and solar wind.

\noindent $\bullet$ Particles with $\Omega_0 \gg \omega$ undergo
strong {\it parallel} heating, qualitatively consistent with linear
theory predictions for the heating produced by the $\mu
\nabla_\parallel B$ forces (``transit time damping'') associated with
the slow magnetosonic waves in MHD turbulence (e.g.,
Fig. \ref{fig:dispersion}, \ref{fig:deltaE}, \ref{fig:scatter}, \&
\ref{fig:Edot}).

\noindent $\bullet$ In contrast to the predictions of linear theory,
we find that discrete resonances {\it do not} dominate the energy
transfer between particles and waves.  Instead, energy transfer occurs
for a wide range of particles, even those that would be quite
``off-resonance'' according to linear theory (Fig. \ref{fig:Landau}).
This is broadly consistent with models in which the rapid
decorrelation of waves in anisotropic MHD turbulence leads to
significant ``resonance broadening'' (e.g., \citealt{chandran2000}).

\noindent $\bullet$ Linear theory predicts that particles with
$\Omega_0 \gg \omega$ undergo purely {parallel} heating because of the
adiabatic invariance of the magnetic moment $\mu \propto \ug^2/B$ in
the presence of low-frequency electromagnetic fluctuations.  Although
we do find that most of the heating is parallel to the local field for
particles with $\Omega_0 \gg \omega$ (Fig. \ref{fig:deltaE} \&
\ref{fig:Edot}) we also see small, but non-zero, changes in $\ug$ and
$\mu$ even for high cyclotron-frequency particles (Fig. \ref{fig:mu}).
The physical origin of this perpendicular heating is not fully clear,
but we speculate that it may be a consequence of resonance broadening
(\S \ref{sec:fiducial}).  We find that the perpendicular heating and
diffusion rates are $\sim 10 \%$ of the parallel heating and diffusion
rates for $\Omega_0 \gg \omega$ (e.g., Fig. \ref{fig:Landau} \&
\ref{fig:Edot}).  Although this is unlikely to change any of the
general conclusions about perpendicular versus parallel heating
derived from linear theory, it is an interesting modification to the
physical picture of how particles interact with anisotropic Alfv\'enic
turbulence.

\noindent $\bullet$ All of the results summarized above are present in
both ideal MHD simulations and in simulations that include an explicit
resistivity to dissipate small-scale fluctuations in the turbulence.
A finite resistivity also generates a non-zero $E_\parallel$ that
produces strong {\it parallel} heating and diffusion of particles
(Fig. \ref{fig:resistive}).  It is important to stress that the
physics of reconnection is not adequately represented by the spatially
and temporally constant resistivity present in our
calculations. Indeed, particle-in-cell and Hall MHD simulations of
reconnection show that particle heating in current sheets is far more
subtle than can be captured by resistive-MHD simulations
\citep{drake2009a,drake2009b}.  Further work is thus required to
understand the particle heating and acceleration produced by current
sheets that self-consistently develop on small scales in a turbulent
plasma.

\noindent $\bullet$ For the particular case in which we initialize a
thermal distribution of test particles having the same sound speed as
the fluid in our turbulence calculations, the test particles
approximate the interaction between protons and the turbulent
fluctuations.  For $\beta \sim 1$, Figure \ref{fig:Edot} shows that
the heating rate of the test particles is $\sim 3 \dot E_{turb}$ for
cyclotron-resonant particles with low $\Omega_0$ and $\sim 0.5 \dot
E_{turb}$ for Landau-resonant particles with high $\Omega_0$ (where
$\dot E_{turb}$ is the energy dissipation rate in the turbulence).
This highlights that in a low-collisionality plasma, the wave-particle
interactions are strong enough to damp a large fraction of the
turbulent energy; the test particle approximation is thus a poor one.

In all of our calculations, the test particles have velocities within
a factor of $\sim 10$ of the sound speed of the fluid $c_s$.  This
corresponds primarily to the velocities of protons or minor ions,
rather than electrons, which have typical velocities $\sim 40 \, c_s$
for $T_e \sim T_p$.  High velocity particles ($\gg c_s$) transit the
computational domain many times in the course of the simulation; doing
so, they may repeatedly sample very similar realizations of the
turbulent fluctuations.  We believe that tests with larger
computational domains are required in order to reliably study the
turbulent diffusion and heating of high velocity particles.  This
limitation has two significant implications.  First, our calculations
do not at this point predict the relative turbulent heating of
electrons and protons, which would be of considerable interest for
both solar and astrophysical problems (e.g.,
\citealt{Quataert:1999,cranmer2009}).  Second, our calculations are
limited in their ability to predict the {\it acceleration} of
particles to energies well above the thermal energy of the plasma.  It
is nonetheless important to highlight that when we initialize a
thermal distribution of test protons that are cyclotron resonant with
the turbulent fluctuations, the distribution function naturally
evolves a non-thermal tail (Fig. \ref{fig:distribution}).

Previous work on test particle heating in MHD turbulence simulations
\citep{dmitruck} concluded that current sheets produce significant
parallel heating of electrons and that ions are preferentially heated
in the perpendicular direction.  Our results are broadly consistent
with these conclusions given Dmitruck et al's assumed particle
gyrofrequencies.  \cite{dmitruck} give an involved physical
interpretation of the perpendicular heating in their calculations; we
suspect that it is simply due to cyclotron resonance, as we have found
in our calculations.\footnote{Although \cite{dmitruck} studied test
  particle heating in a static snapshot of MHD turbulence (i.e.,
  $\omega = 0$) their particles could nonetheless undergo cyclotron
  resonance via the parallel Doppler shift in eq. (\ref{eq:res}).}
Perhaps most importantly, the perpendicular ion heating found both
here and in \cite{dmitruck} is not, we believe, applicable to the
solar wind, contrary to the claims made by Dmitruck et al.  In
particular, as we now explain, \cite{dmitruck} did not consider the
limitations imposed by finite numerical resolution when claiming that
their results could be directly applied to the solar wind.

%\vspace{0.2cm}

\subsection{Implications}

\label{sec:implication}

Broadly speaking, our results suggest that many of the conclusions
drawn from quasilinear theory about the heating and acceleration of
particles by anisotropic Alfv\'enic turbulence are likely to be
qualitatively correct.  However, many of the quantitative results may
change given, e.g., the lack of the discrete resonances that strongly
shape the predictions of quasilinear theory
(Fig. \ref{fig:Landau}). Assessing in detail the implications of our
results for heating by MHD turbulence in solar and astrophysical
environments will require additional work.  Here we take the
near-Earth solar wind as an example to illustrate the implications of
our results and the limitations due to finite numerical resolution.

A number of observations indicate that the solar wind undergoes
spatially extended heating.  For example, {\em in situ} measurements
from satellites such as Helios \& Ulysses show that electrons and
protons have non-adiabatic temperature profiles (e.g.,
\citealt{cranmer2009}).  Heating by anisotropic Alfv\'enic turbulence
is one of the leading models for the origin of this non-adiabiticity
(e.g., \citealt{matthaeus1999,cranmer2003}).  {\em In situ}
measurements of the solar wind at 1 AU show that the proton
distribution function is on average anisotropic with respect to the
local magnetic field: $T_\perp \simeq 0.9 \, T_\parallel$
\citep{bale2009}, although the sign of the anisotropy depends on the
wind speed, with $T_\perp \gtrsim T_\parallel$ for $v_{wind} \gtrsim
600$ km s$^{-1}$ and $T_\perp \lesssim T_\parallel$ for $v_{wind}
\lesssim 600$ km s$^{-1}$ \citep{kasper2002,hellinger2006}.  It is not
yet clear how to understand these measured temperature anisotropies in
the context of heating by anisotropic Alfv\'enic turbulence.

Typical physical parameters for the slow solar wind\footnote{Our
  simulations have roughly the same energy flux traveling in both
  directions along the local magnetic field.  This is typically not
  true at $\sim 1$ AU in the solar wind, particularly in the fast wind
  \citep{marsch2006}.  For this reason, we compare to measurements in
  the slow solar wind, where the assumption of balanced turbulence is
  more consistent with the measurements.}  at $\sim 1$ AU near Earth
are $B \simeq 10^{-4}$ G, $n_0 \simeq 20$ cm$^{-3}$, $T_p \simeq T_e
\simeq 1.5 \times 10^5$ K, $v_A \simeq 50$ km s$^{-1}$, $c_s \simeq
35$ km s$^{-1}$, $v_{\rm wind} \simeq 460$ km s$^{-1}$, and $\beta
\simeq 0.4$ (e.g., \citealt{celnikier1987}); the proton Larmor radius
and gyrofrequency are thus $r_{L,p} \simeq 3 \times 10^{6}$ cm and
$\Omega_{p} \simeq 0.15$ Hz, respectively, while the electron Larmor
radius and gyrofrequency are $r_{L,e} \simeq 10^{5}$ cm and
$\Omega_{e} \simeq 300$ Hz, respectively.  \cite{howes2008a} reviewed
a variety of observational diagnostics of the outer scale of the
turbulence in the solar wind and estimated that $k_{min}^{-1} \sim
10^{10-11}$ cm.  The cyclotron frequency in our calculations is
expressed in units of $c_s/L$ where the size of our box $L \sim 2
\pi/k_{min}$ is also of order the outer-scale of the turbulence
$k_{min}^{-1}$.  Expressed in these units, the proton and electron
cyclotron frequencies in the solar wind are $\simeq 10^{4-5} \, c_s/L$
and $\simeq 10^{7-8} \, c_s/L$, respectively.  The {\it minimum} value
of the proton cyclotron frequency in the solar wind is thus comparable
to the {\it maximum} value of the cyclotron frequency of particles
that we have simulated (e.g., Fig \ref{fig:Edot}).  The fundamental
reason for this is that, in nearly all heliospheric and astrophysical
plasmas, the ratio of the proton cyclotron frequency to the
outer-scale frequency of MHD turbulence is much larger than the
dynamic range that can be simulated with current computational
resources.

A naive application of our results to the near-Earth solar wind, using
the parameters above, suggests that all particle heating by MHD
turbulence corresponds to Landau-resonant particles (high $\Omega_0$)
in the terminology of this paper.  This is indeed correct, {for the
  outer-scale turbulent fluctuations that we can simulate.}  If a slow
wave cascade is present at $\sim 1$ AU, our $\beta = 1$ simulations in
Figure \ref{fig:Edot} demonstrate that most of the slow-wave energy on
large-scales is likely lost to particle heating, in particular
parallel heating of the protons.  In addition to being important for
the thermodynamics of the solar wind, the slow waves are one of the
primary sources of density fluctuations in anisotropic Alfv\'enic
turbulence \citep{lg2001}; if they are indeed largely damped at large
scales when $\beta \sim 1$, this suppresses the contribution of slow
waves to the small-scale density fluctuations that are directly
measured in the solar wind (see \citealt{chandran2009}).

Although the slow wave component of the cascade can lose a significant
fraction of its energy on large scales to $\mu \nabla_\parallel B$
acceleration of particles (Fig. \ref{fig:Edot}), the Alfv\'en-wave
component of the cascade does not produce significant particle heating
until (1) the Alfv\'en wave frequency becomes comparable to the
cyclotron frequency of particles that have a significant density
(e.g., protons or helium) or (2) the {\it perpendicular} wavelength of
the Alfv\'en waves becomes comparable to the proton Larmor radius
$r_{L,p}$, at which point the Alfv\'en wave cascade transitions to a
kinetic Alfv\'en wave cascade \citep{howes2008a}.

From Figure \ref{fig:max} we conclude that the maximum gyrofrequency
of particles that can be cyclotron-resonant with the turbulence is
$\Omega_{max} \simeq 2 \, k_{\perp,max}^{2/3} k_{min}^{1/3} v_A$,
where we have identified $k_{max} \simeq k_{\perp, max}$ given the
anisotropy of the Alfv\'enic cascade.  Using the parameters for the
solar wind above, we estimate that the maximum gyrofrequency of a
particle that can be cyclotron-resonant when the cascade reaches
$k_{\perp, max} \sim r_{L,p}^{-1}$ is $\Omega_{max} \simeq 0.02$ Hz
$\ll \Omega_p \simeq 0.15$ Hz. As a result direct cyclotron resonance
{\it is not} important at $k_{\perp, max} \sim r_{L,p}^{-1}$; instead,
the dissipation of the Alfv\'enic component of weakly compressible MHD
turbulence occurs via the kinetic Alfv\'en wave cascade launched when
$k_{\perp, max} \sim r_{L,p}^{-1}$.  Our results based on the direct
integration of particle orbits in fully developed MHD turbulence
support previous work that reached the same conclusion using linear
theory cascade models (e.g.,
\citealt{Quataert:1999,cranmer2003,howes2008a}).  

The dissipation of anisotropic kinetic Alfv\'en wave turbulence having
$k_{\perp} \gtrsim r_{L,p}^{-1}$ is still not fully understood;
possibilities include Landau resonance \citep{howes2008a}, secondary
instabilities that generate cyclotron frequency waves
\citep{markovskii2006}, dissipation in current sheets
\citep{drake2009b}, and stochastic ion orbits created by sufficiently
large amplitude waves \citep{johnson2001,voitenko2004}.  Our results,
together with the empirical evidence for energetically important
kinetic Alfv\'en waves in the near-Earth solar wind (e.g.,
\citealt{sahraoui2009}), strongly suggest that the anisotropic proton
and ion temperatures in the solar wind are in part a consequence of
heating by this kinetic Alfv\'en wave cascade.

\acknowledgements We thank Xuening Bai, Ben Chandran, Jim Drake, Jeff
Oishi, and Jim Stone for useful discussions.  The Athena resistivity
module was provided by Jim Stone.  Computing time was provided by the
National Science Foundation through the Teragrid resources located at
the National Center for Atmospheric Research and the Pittsburgh
Supercomputing Center.  Support for I.~J.~.P was provided by NASA
through Chandra Postdoctoral Fellowship grant PF7-80049, awarded by
the Chandra X-Ray Center, which is operated by the Smithsonian
Astrophysical Observatory for NASA under contract NAS8-03060.
E.~Q. and R.~L. were supported in part by NSF-DOE Grant PHY-0812811,
NSF Grant ATM-0752503, and by the David and Lucille Packard
Foundation.  EQ was also supported in part by the Miller Institute for
Basic Research in Science, University of California Berkeley. \\

\bibliography{references}

\appendix

\section{Interaction of particles with the turbulent
  driving}\label{ap:driving}

\begin{figure} 
\centering
\includegraphics[scale=0.45]{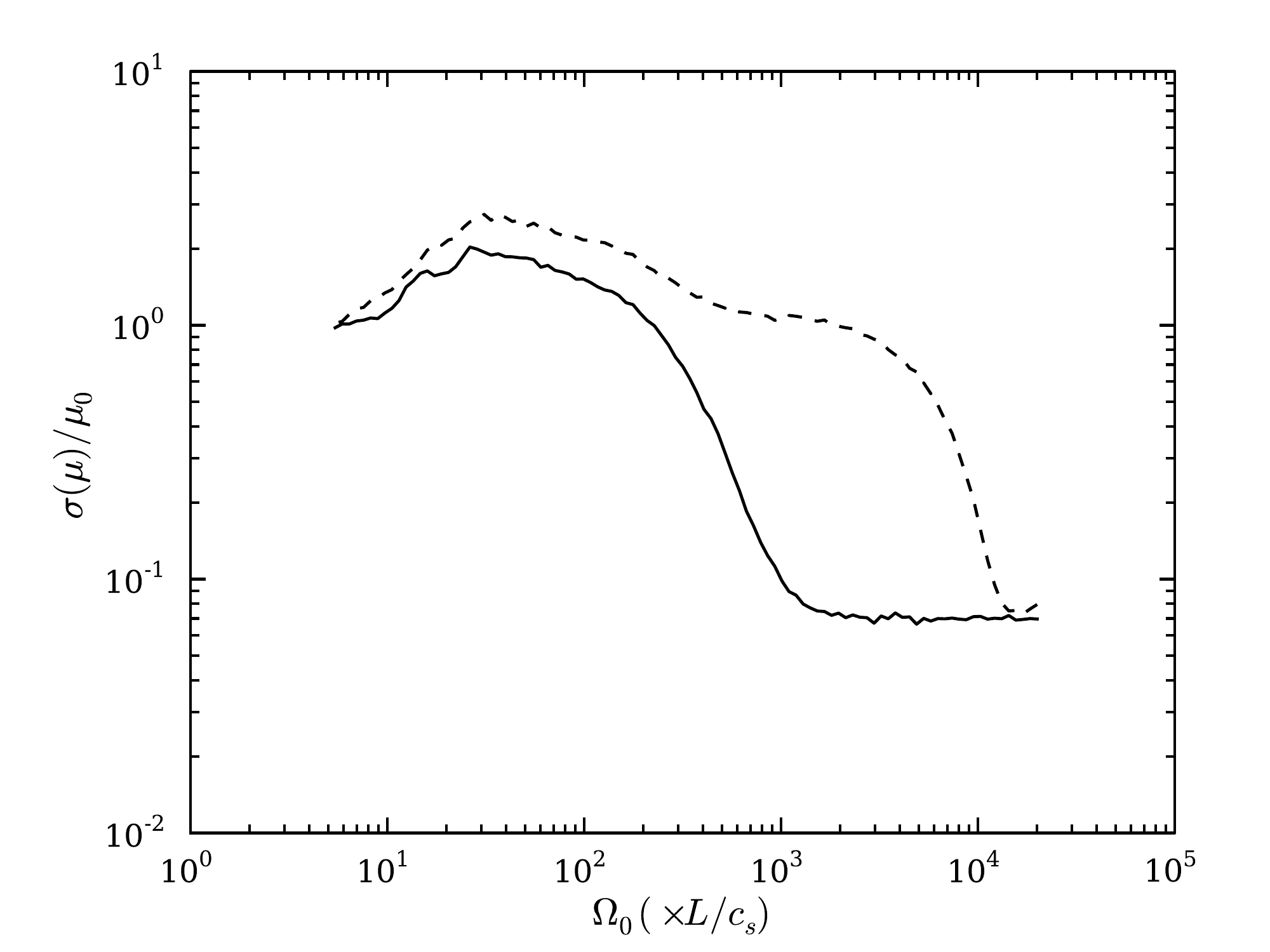}
\caption[Variations of $\mu$ induced by the turbulent
driving]{Standard deviation of the magnetic moment $\mu$ as a function
  of the particles' gyrofrequency $\Omega_0$, after an integration
  time of $\simeq L/c_s$. After the particle integration begins we
  either let the turbulence decay (solid line) or keep driving it
  (dashed line). The parameters of the simulation are summarized in
  Table \ref{tab:fiducial}.  The continued driving of the turbulence
  (dashed line) introduces spurious high frequencies into the problem
  which artificially increase the magnetic moment at high $\Omega_0$.}
\label{fig:driving}
\end{figure}

As mentioned in \S \ref{sec:integration}, driving the turbulence can
have undesirable effects on the particles' motion. Our default method
for driving the turbulence introduces random kicks to the velocity
field at each MHD timestep. This artificially introduces variations
that are much faster than the MHD evolution, and potentially faster
than the particles' gyration. As a result, the conservation of the
magnetic moment can be artificially modified.  Figure
\ref{fig:driving} compares the standard deviation of the magnetic
moment in two simulations: we either continue driving the turbulence
after the particle integration is turned on (dashed) or let the
turbulence decay (solid). The particles initially have $\uo = 1.0 \,
c_s$ and $\mu_0 = 1.0 \, c_s^2/B_0$, with no variance in $\mu$. After
an integration time $\simeq L/c_s$, the two simulations produce quite
different dispersions in $\mu$, in particular at high $\Omega_0$,
where the gyrofrequency is much larger than the true frequencies of
the fluctuations resolved in the simulation.  The continued driving
introduces spurious high frequencies into the problem, which
significantly change the resulting particle heating.  This can be
remedied by driving with a decorrelation time comparable to the
outer-scale turnover time of the turbulence, but we choose to be
conservative and study the particle heating in decaying turbulence.

\section{Tests}\label{ap:tests}

After writing the particle integrator, we verified it with a series of
tests. We ran it with several simple field configurations and checked
that we recovered known solutions. We also studied how the results
depended on the chosen time step. Here we summarize some of our tests
and their results.

\subsection{Constant magnetic field}

The simplest test is to confirm that one obtains the well-known
helical motion in the case of a constant and uniform magnetic field
and no electric field.  For this problem, our method conserves the
particles' energy and magnetic moment to machine accuracy. The Boris
algorithm tends to produce a slightly low gyrofrequency and a slightly
large gyroradius. However, with our choice of 40 time steps per
gyration, the relative error is $\simeq 10^{-3}$.  Note also that this
simply produces a systematic change in the definition of the charge to
mass ratio of the particles; there is no secular change in time.

\subsection{Alfv\'en wave}

\begin{figure} 
\centering
\includegraphics[scale=0.45]{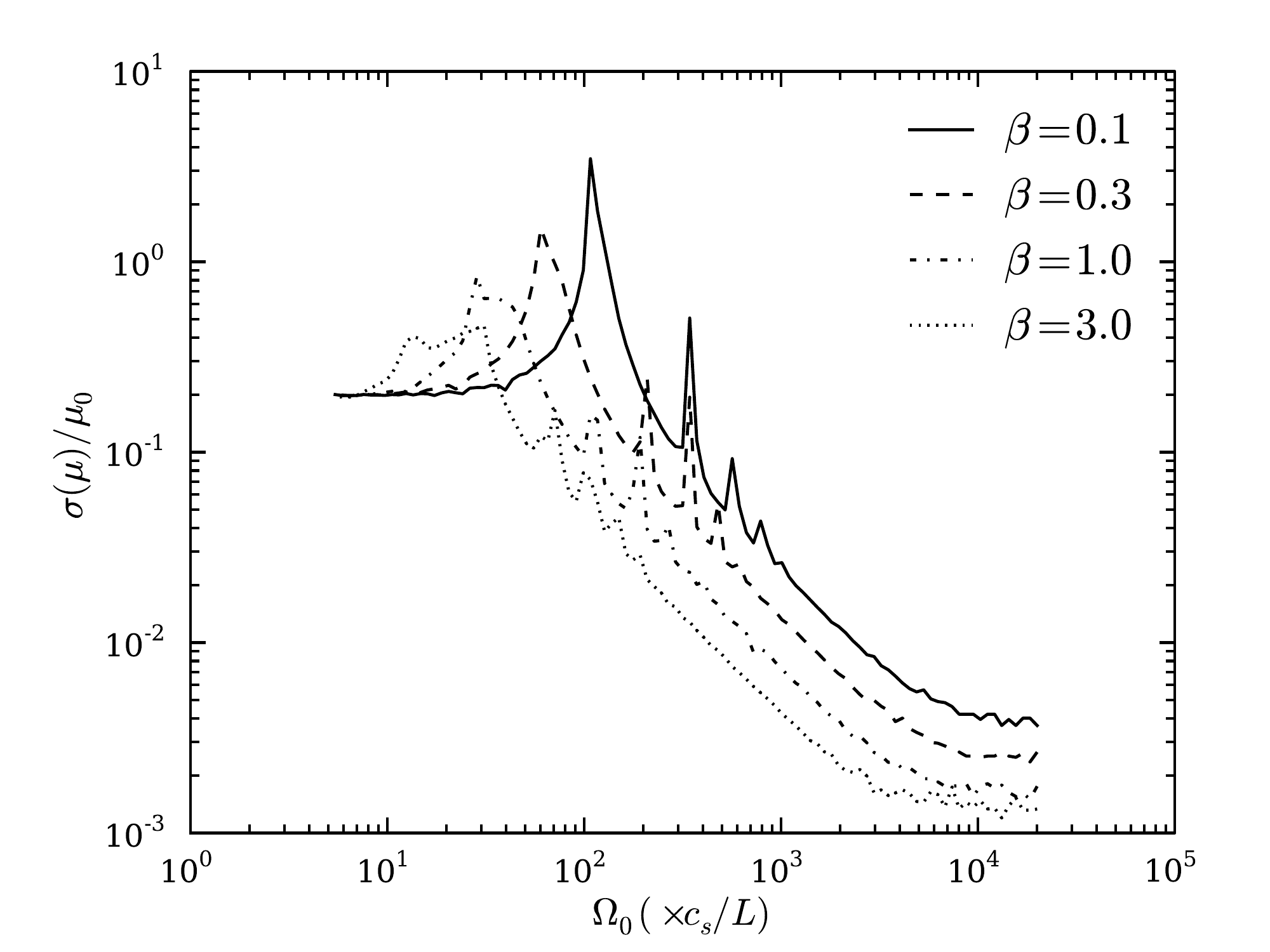}
\caption[Dispersion in $\mu$ due to an Alfv\'en wave]{Standard
  deviation of the magnetic moment $\mu$ as a function of the
  particles' gyrofrequency $\Omega_0$, due to interaction with a
  single parallel-propagating Alfv\'en wave, after an integration time
  of $\simeq L/c_s$. The particles' initial velocities are $\uo = 0$
  and $\ugo = 1.0\,c_s$. The curves correspond to different values of
  $\beta$, but in all cases the wave has the same amplitude ($\delta
  v= 0.1\,c_s$) and the same wavenumber ($ k = 4 \times 2\pi/L$).}
\label{fig:wave}
\end{figure}

We let the particles evolve in the presence of a single Alfv\'en wave
propagating parallel to the background magnetic field. The particles'
velocities are initialized according to the delta function $f_{0}$
from equation (\ref{eq:init}), so they all start with the same
magnetic moment. Figure \ref{fig:wave} shows $\sigma(\mu)$ after an
integration time of $\sim L/c_s$ for different values of $\beta$ (and
thus different values of $v_{A}$).  The magnetic moment is conserved
to high accuracy for particles with gyrofrequencies much larger than
the frequency of the wave. We also observe several resonances that
produce large changes in $\mu$.  These are consistent with cyclotron
resonances, and their positions scale as $\beta^{-1/2}$, as predicted
by linear theory.

\begin{figure} 
\centering
\includegraphics[scale=0.45]{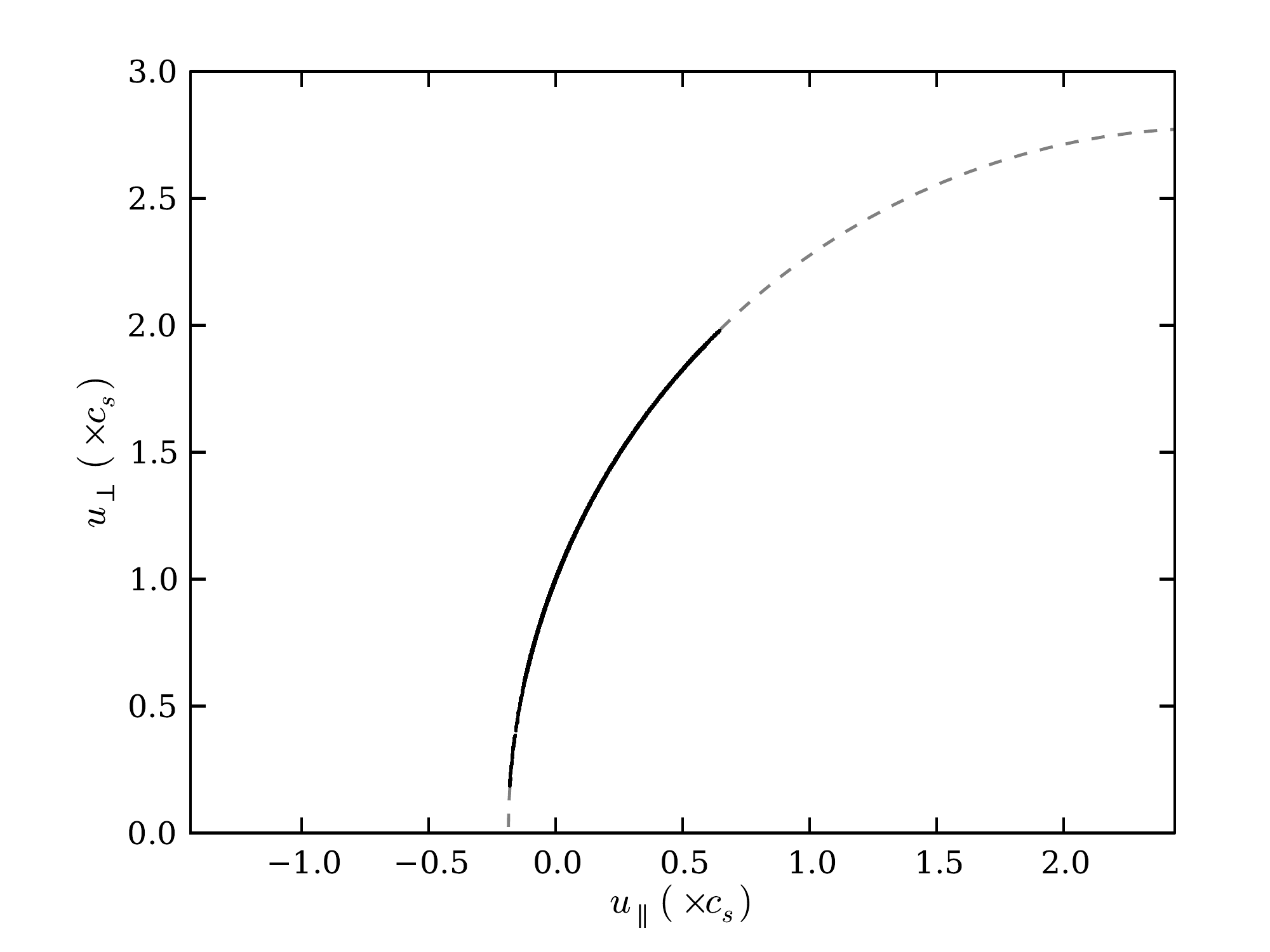}
\caption[Scatter plot under an Alfv\'en wave]{Scatter plot in velocity
  space ($\upa$,$\ug$), for cyclotron-resonant particles ($\Omega_0 =
  60\, c_s/L $) interacting with a plane Alfv\'en wave ($\delta v =
  0.1 \, c_s$; $k= 4 \times 2\pi/L $) at $\beta = 0.3$.  The gray
  dashed circle has its center at $(v_A,0)$ -- with $v_A = 2.6\,c_s$
  -- and represents constant energy in the frame of the wave.  The
  particles initially all have the same values of $\upa$ and $\mu$;
  the black dots show the location of the particles after interacting
  with the wave for a time $\simeq L/c_s$.  As required analytically,
  the particles move only along the gray dashed curve.}
\label{fig:circle}
\end{figure}

An important feature of Alfv\'en waves is that the perturbed electric
field vanishes \emph{in the frame of the wave}. As a consequence, the
energy of a particle cannot change in this frame. Back in the frame of
the fluid: \be u_{\perp}^2 + (u_{\parallel} - v_{A})^2 = \epsilon \ee
where $\epsilon$ is the initial energy of the particle. Particles thus
evolve on a sphere in velocity space. Figure \ref{fig:circle} shows
how cyclotron-resonant particles are scattered by an Alfv\'en wave in
velocity space.  The wave has an amplitude $\delta v = 0.1\, c_s$
 and wavenumber $k = 4 \times 2\pi/L$; $\beta = 0.3$.  The
fact that particles remain on a circle centered on $v_{A}$ confirms
that our integrator accurately reproduces the properties of the
resonance.

\subsection{Particle's timestep}

As explained in \S \ref{sec:integration}, for particles with
normalized gyrofrequency $\Omega_0$, our integration time step is
chosen to be \be \Delta t_{particle} = \mathrm{min} \left(
  \frac{1}{N_1} \frac{2\pi}{\Omega_0} \, , \, \frac{1}{N_2} \Delta
  t_{MHD} \right) \ee where $N_1$ corresponds to the minimal number of
particle time steps per gyration, and $N_2$ corresponds to the minimal
number of particle time steps within one MHD time step.  In our
simulations, we use $N_1 = 40$ and $N_2 = 10$. For $N_2 \gtrsim 3$,
there was no significant change in the results.  Similarly, so long as
$N_1 \gtrsim 10$, we found that the results were converged to better
than $\sim 1 \%$.

\section{Interpolation Method}\label{ap:interpolation}

This section gives more details on how we interpolate the fields
$\bs{E}$ and $\bs{B}$ from the MHD grid to the particle's position. We
use a directionally split interpolation algorithm.  The fields are
averaged with weights that are the product of four 1-dimensional
weights. If $(x,y,z,t)$ is the current position of the particle,
$i_0$, $j_0$, $k_0$ the indices of the nearest cell, and $l_0$ the
index of the nearest MHD timestep: \be \overline{\bs{F}} =
\sum_{i=i_0-1}^{i_0+1} \sum_{j=j_0-1}^{j_0+1} \sum_{k=k_0-1}^{k_0+1}
\sum_{l=l_0-1}^{l_0+1} w_{i}(x) \: w_{j}(y) \: w_{k}(z) \: w_{l}(t) \,
\bs{F}_{i,j,k,l} \ee where the $\bs{F}$ stands for $\bs{E}$ or
$\bs{B}$, and the overline represents interpolation.

Choosing appropriate 1-D weights is crucial. In particular, ill-chosen
expressions may lead to discontinuities in the interpolated fields as
particles cross from one cell to another, and have proved to produce
spurious variation in the particles' energy. As explained in \S
\ref{sec:interpolation}, we use the Triangular Shaped Cloud (TSC)
method, which ensures that the interpolated fields have
$\mathcal{C}^{1}$ smoothness in both space and time. In the $x$ direction, for instance:
\begin{align}
w_{i_0-1}(x) & = \frac{1}{2} \left( \frac{1}{2} - \frac{(x-x_{i_0})}{\Delta x} \right)^2 \\
w_{i_0}(x) & = \frac{3}{4} - \left( \frac{x-x_{i_0}}{\Delta x} \right)^2 \\
w_{i_0+1}(x) & = \frac{1}{2} \left( \frac{1}{2} + \frac{(x-x_{i_0})}{\Delta x} \right)^2
\end{align}
where $x_{i_0}$ is the $x$ position of the nearest cell's center and $\Delta x$ is the grid spacing. Weights in the $y$, $z$ and $t$ directions have similar expressions.

\end{document}